\documentclass[aps,prd,twocolumn,showpacs,superscriptaddress,preprintnumbers,nofootinbib,
amsmath,amssymb]{revtex4-1}

\usepackage{graphicx} 
\usepackage{tikz}
\usetikzlibrary{arrows}
\usepackage{multirow}
\usepackage[colorlinks=true,urlcolor=blue,linkcolor=blue,citecolor=blue]{hyperref}
\usepackage[normalem]{ulem}
\usepackage[capitalize]{cleveref}
\usepackage{amsfonts,amssymb}
\usepackage{comment}
\usepackage{flushend}
\usepackage[T1]{fontenc}
\usepackage{mathtools}
\usepackage{xcolor}
\usepackage{graphicx}
\usepackage{dcolumn}
\usepackage{bm}

\begin{document}
\preprint{KEK-QUP-2025-0012, KEK-TH-2726, TU-1264}

\title{Light Dark Matter Detection with Sub-eV Transition-Edge Sensors} 

\author{Muping Chen}
\email{mpchen@post.kek.jp}
\affiliation{International Center for Quantum-field Measurement Systems for Studies of the Universe and Particles (QUP, WPI),
High Energy Accelerator Research Organization (KEK), Oho 1-1, Tsukuba, Ibaraki 305-0801, Japan}

\author{Volodymyr Takhistov}
\email{vtakhist@post.kek.jp}
\affiliation{International Center for Quantum-field Measurement Systems for Studies of the Universe and Particles (QUP, WPI),
High Energy Accelerator Research Organization (KEK), Oho 1-1, Tsukuba, Ibaraki 305-0801, Japan}
\affiliation{Theory Center, Institute of Particle and Nuclear Studies (IPNS), High Energy Accelerator Research Organization (KEK), 1-1 Oho, Tsukuba, Ibaraki 305-0801, Japan}
\affiliation{Graduate University for Advanced Studies (SOKENDAI), 1-1 Oho, Tsukuba, Ibaraki 305-0801, Japan}
\affiliation{Kavli Institute for the Physics and Mathematics of the Universe (WPI), UTIAS, The University of Tokyo, Kashiwa, Chiba 277-8583, Japan}

\author{Kazunori Nakayama}
\email{kazunori.nakayama.d3@tohoku.ac.jp}
\affiliation{Department of Physics, Tohoku University, Sendai, Miyagi 980-8578, Japan
}
\affiliation{International Center for Quantum-field Measurement Systems for Studies of the Universe and Particles (QUP, WPI),
High Energy Accelerator Research Organization (KEK), Oho 1-1, Tsukuba, Ibaraki 305-0801, Japan}

\author{Kaori Hattori}
\email{hattorik@post.kek.jp}
\affiliation{International Center for Quantum-field Measurement Systems for Studies of the Universe and Particles (QUP, WPI),
High Energy Accelerator Research Organization (KEK), Oho 1-1, Tsukuba, Ibaraki 305-0801, Japan}

 \begin{abstract}
We present a comprehensive analysis of high-resolution transition-edge sensors (TESs) as a quantum sensing platform for detecting dark matter (DM). Operating near the thermodynamic noise limit with sub-eV energy resolution, TESs offer a powerful approach for probing light DM in the sub-GeV mass range. Optical TESs, realized on superconducting films with critical temperatures below 150~mK, achieve energy thresholds below 100~meV and enable precise calorimetric detection of individual energy depositions. We model TES response by incorporating fundamental noise sources and applying optimal filtering techniques, and evaluate their sensitivity across a range of DM interaction channels, accounting for in-medium effects in the target material. We show that even ng-month-scale exposures can reach previously unexplored DM–electron scattering cross sections below $10^{-27}~\mathrm{cm}^2$ for sub-MeV masses, and can probe the MeV-scale mass range for DM–nucleon couplings.  Combining high energy resolution, photon-number sensitivity, and scalability, optical TESs provide a compelling quantum sensing platform for rare-event searches at the intersection of particle physics and quantum metrology.
 \end{abstract}

\maketitle

\section{Introduction}

Dark matter (DM), which constitutes approximately $85\%$ of the matter content of the Universe, remains one of the most significant open problems in physics. While decades of experimental efforts have focused on weakly interacting massive particles (WIMPs) with masses above the GeV scale, conclusive detection observations are absent. There is a rapidly growing interest and frontier aiming to explore sub-GeV DM candidates~\cite{Essig:2022dfa}. 
Detection of light DM through its scattering/absorption by electrons/phonons in various type of materials have been extensively studied in Refs.~\cite{Essig:2011nj,Essig:2015cda,Hochberg:2015fth,Hochberg:2016ajh,Hochberg:2016sqx,Hochberg:2017wce,Knapen:2017ekk,Griffin:2018bjn,Trickle:2019nya,Gelmini:2020xir,Chen:2021qao,Hochberg:2021pkt,Knapen:2021run,Knapen:2021bwg,Mitridate:2021ctr,Campbell-Deem:2022fqm,Trickle:2022fwt,Chen:2022xzi,Mitridate:2023izi,Griffin:2024cew}.
These lighter DM particles typically deposit only $\sim$eV-scale energies in interactions, pushing detection requirements into new regimes. Realizing sensitivity to such low-energy deposits calls for quantum sensors with single-quanta resolution, sub-eV thresholds, and low noise levels. Recently, several classes of superconducting sensors have been proposed for DM detection, including superconducting nanowire single photon detector (SNSPD)~\cite{Hochberg:2019cyy,Hochberg:2021yud,Chiles:2021gxk} and kinetic inductance detector (KID)~\cite{Gao:2024irf} as both target materials and sensors for DM detection. Here, we investigate and demonstrate the distinct potential of sub-eV resolution superconducting transition-edge sensors (TESs) for highly sensitive DM detection.

Superconducting TESs~\cite{Irwin2005} constitute a broad class of quantum detectors that are especially well-positioned to fulfill stringent requirements to realize necessary DM sensitivity. Operated at the sharp superconducting-normal transition, TESs transduce minute energy deposits into detectable resistance changes via strong electrothermal feedback. A TES ``thermometer''
can be utilized either as a bolometer to measure continuous power input or a calorimeter to measure energy depositions. TES-based detectors have already demonstrated their potential across a range of basic physics applications.
Notably, the TESSERACT experiment recently reported the first TES-based nuclear recoil limits on DM as light as $\sim$50 MeV, using silicon targets with thresholds in the 10-100 eV range~\cite{TESSERACT:2025tfw}. TESs have also been proposed or implemented in diverse experimental contexts related to physical processes, including relic neutrino detection~\cite{PTOLEMY:2018jst}, coherent elastic neutrino-nucleus scattering~\cite{Ricochet:2023yek}, neutrinoless double beta decay~\cite{Bratrud:2024fmj}, and calorimetric electron-capture measurements~\cite{Alpert:2025tqq}. Ongoing developments aim to reduce photon-like dark counts to levels compatible with rare-event searches~\cite{Manenti:2024etv}.

Optical TESs have emerged as a highly sensitive TES class with the ability to resolve individual eV-scale photons with exceptional precision. These devices operate near the boundary of classical and quantum thermodynamics, converting the energy of a single photon into a quantifiable signal. Recently, optical TESs using ultra-thin Au/Ti bilayers with critical temperatures below 120 mK have achieved full-width-at-half-maximum (FWHM) energy resolutions as low as $\Delta E \simeq 67$ meV at 0.8 eV~\cite{Hattori:2022mze}. This performance places highly sensitive optical TESs in the quantum calorimetric regime, where thermodynamic energy fluctuations are governed by quantum statistics and individual energy quanta become resolvable. Optical TESs are qualitatively distinct from the calorimetric TESs typically used in x-ray spectroscopy. They offer not only superior energy resolution, but also faster response times and the potential for scalable array integration.

In addition to their high energy resolution, optical TESs offer true photon-number-resolving capability. When illuminated by a monochromatic light source, the number of simultaneously absorbed photons can be determined from the total deposited energy. This property can be  crucial for quantum optics and quantum information applications, where distinguishing between photon events is essential. Photon-number resolution has enabled key advances in quantum computing~\cite{Knill:2001lrt}, quantum communication and receivers~\cite{Becerra:2015}, and quantum metrology~\cite{Matthews:2016,vonHelversen:2019}. Combined with low dark count rates and sub-eV thresholds, this functionality makes optical TESs stand out among quantum detectors able to provide both spectral and number-resolving information, with applications spanning from quantum state reconstruction to multiplexed imaging and fundamental physics searches.
Here, we demonstrate that high resolution optical TESs offer a powerful platform for DM detection, with the ability to probe rare energy depositions from DM across a wide range of interaction types. Their sub-eV sensitivity, combined with quantum-limited resolution and well-understood thermal modeling, opens new discovery space for low-mass DM.

In this work we present the first comprehensive end-to-end modeling of optical TES performance in the sub-eV energy regime and demonstrate their unprecedented capabilities for DM detection. Our analysis includes detailed treatments of TES noise sources, optimal signal filtering, and full in-medium interaction calculations across a range of DM detection channels. While other quantum-sensing efforts, such as ALPS II, are investigating TESs for photon detection in axion-like particle searches~\cite{Schwemmbauer:2024jel}, our approach distinctly combines ultra-low TES thresholds, material-specific response modeling, and realistic detector simulations to project sensitivities for next-generation dark matter experiments. These results show that optical TES arrays are not only viable, but also highly competitive quantum sensors for broad rare-event searches at the intersection of particle physics and quantum metrology.

In Sec.\ref{sec:highres}, we characterize the basic properties of high-resolution optical TESs and perform simulations and fits of their response. In Sec.~\ref{subsec:AlResponse}, we discuss key quantities for characterizing the target material, including the dynamic structure factor and the energy loss function. In Sec.\ref{sec:dme}, we derive sensitivity projections for various DM search channels, including DM–electron scattering, DM–nucleon scattering, DM absorption, the Migdal effect, and power deposition. We conclude in Sec.\ref{sec:conc}.

\section{High Resolution Optical TES}
\label{sec:highres}

\subsection{Optical TES characterization}

\begin{figure}[t]
    \centering
    \includegraphics[width=0.7\linewidth]{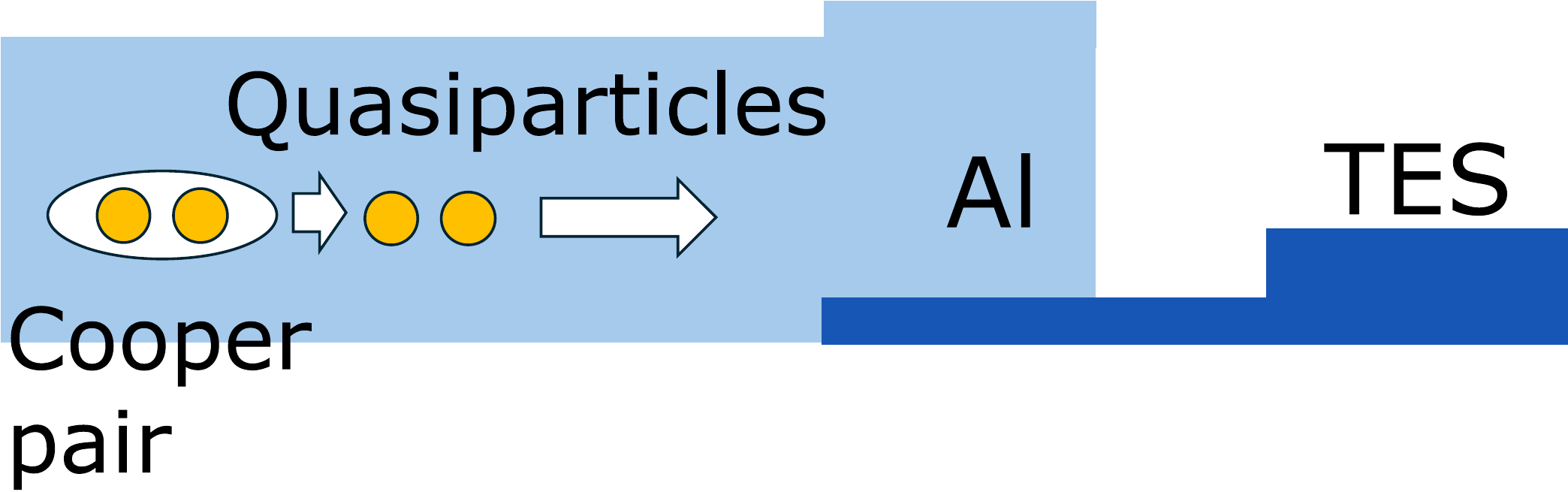}
    \caption{Setup for quasiparticle-trap-assisted electrothermal-feedback
transition-edge sensors (QETs).}
    \label{fig:schematicTES}
\end{figure}

\begin{figure*}[t]
    \centering
\includegraphics[width=0.35\linewidth]{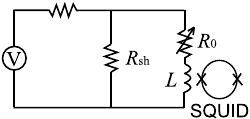} \hspace{4em}
\includegraphics[width=0.35\linewidth]{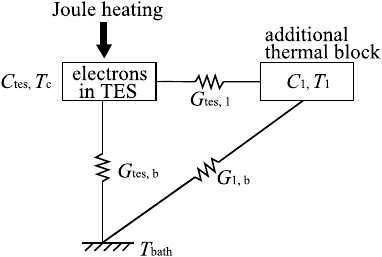}
    \caption{[Left] Bias circuit of TES-R. [Right] Thermal model of the TES-R.}
    \label{fig:TES-model}
\end{figure*}

As a representative high resolution TES   for DM searches, we consider an 
optical TES that is
$20$~nm/10~nm Ti/Au bilayer with critical temperature of $T_{c}=115$~mK, as recently realized in Ref.~\cite{Hattori:2022mze}.
Fig.~\ref{fig:schematicTES} illustrates a schematic of the detection concept. DM interactions deposit energy into a superconducting target material (e.g. Al), breaking Cooper pairs and generating quasiparticles. These excitations are subsequently thermalized and measured by the quasiparticle-trap-assisted electrothermal-feedback transition-edge sensor (QET) ~\cite{Irwin:1995qet}, which operates near the superconducting transition to achieve high energy resolution.
The TES operates near the fundamental noise limit, where the dominant noise contributions arise from intrinsic thermal and Johnson fluctuations. These limits define the ultimate sensitivity achievable for energy deposition events, placing optical TESs among the most sensitive quantum calorimeters currently realizable.

In Fig.~\ref{fig:TES-model} we depict 
bias circuit as well as thermal model of TES. Here, voltage bias is supplied through a shunt resistor of $R_{\rm sh} = 18$~m$\Omega$. The TES resistance changes are converted into measurable current signals, which are read out using a superconducting quantum interference device (SQUID) of input inductance $L = 18$~nH. The SQUID readout provides sub-eV energy sensitivity with minimal excess noise.
For characterization of TES properties and behavior we follow Ref.~\cite{Irwin2005}.

In Tab.~\ref{tableTES} we summarize the relevant electrical and thermal parameters of reference TES (TES-R) we consider at the bias point
$R/R_{n}=0.10$, where $R$ is the TES resistance and $R_n$ is its normal state value.  
The steady-state bias current is $I_0$ and the TES resistance at that point is $R_0$.  
Logarithmic temperature and current sensitivities are
\begin{equation}
\alpha = \dfrac{T}{R_0}\dfrac{\partial R}{\partial T}~~~,~~~
 \beta = \dfrac{I_0}{R_0}\dfrac{\partial R}{\partial I}\Big|_T~.  
\end{equation}   
Heat flows from the TES electron system to the bath through two thermal conductances $G_{\mathrm{tes,b}}$ (TES–bath) and through $G_{\mathrm{tes,1}}$ (TES–absorber node $C_1$).  
The effective small-signal thermal conductance is
$
  G_{\mathrm{eff}} = G_{\mathrm{tes,b}} + G_{\mathrm{tes,1}}~.
$
The associated loop gain 
\begin{equation}
  \mathcal{L}_{\mathrm{eff}}
  = \dfrac{I_0 R_0 \alpha}{G_{\mathrm{eff}} T_c}
  \simeq 53
\end{equation}
is sizable, ensuring strong electro-thermal feedback (ETF) and fast return to equilibrium recovery time.

The small-signal thermal time constants are
\begin{equation}
  \tau_0
  = \dfrac{C_{\mathrm{tes}}}{G_{\mathrm{tes,b}}+G_{\mathrm{tes,1}}} ~~~,~~~ \tau_1
  = \dfrac{C_{1}}{G_{\mathrm{tes,1}}}
\end{equation}
while ETF modifies the electrical time constant to
$
  \tau_I = \tau_0 / (1-\mathcal{L}_{\mathrm{eff}}).
$
Coupling between the two thermal nodes is described by
\begin{equation}
  g_{\mathrm{2\text{-}body}}
  =
  \frac{G_{\mathrm{tes,1}}(T_{\mathrm{tes}}) G_{\mathrm{tes,1}}(T_{1})}
       {(G_{\mathrm{tes,1}}(T_{\mathrm{tes}})+G_{\mathrm{tes,b}})
        \bigl(G_{\mathrm{tes,1}}(T_{1})+G_{1,\mathrm{b}}\bigr)}.
\end{equation}
where $T_{\mathrm{tes}}$ and $T_1$ are the electron temperatures of the TES and node $C_1$, respectively.
 
The complex TES impedance  is then determined entirely by the parameters in
Tab.~\ref{tableTES} as
\begin{align}
\label{eq:impedance}
Z_{\rm TES}(\omega)=&~R_{0}(1+\beta) +
\frac{R_{0}\mathcal{L}_{\mathrm{eff}}(2+\beta)}{1-\mathcal{L}_{\mathrm{eff}}} \\
&~ \times\left[
 1+i\omega\tau_{I}-\frac{g_{\mathrm{2-body}}}
          {(1-\mathcal{L}_{\mathrm{eff}}) \left(1+i\omega\tau_{1}\right)}
\right]^{-1}. \notag
\end{align} 
This impedance is the main electrical input required for the signal characterization we describe below.

Two intrinsic fundamental noise sources dominate in the 100 Hz-1 MHz band and set the ultimate sensitivity of TES, Johnson noise and thermal-fluctuation noise (TFN). The Johnson current noise given by
\begin{equation}
\label{eq:Jnoise}
|I_{\omega}|_{\mathrm{J}}^{2}
=4k_{B}T_cI_0^2R_0\frac{(1+M^2)}{\mathcal{L_{\rm eff}}}(1+\omega^2\tau^2_0)|s_I(\omega)|^2~,
\end{equation}
where $M$ is empirically obtained from data to determine excess Johnson noise~\cite{Hattori:2022mze}.
The small-signal responsivity is given by
\begin{eqnarray}
    s_I(\omega)=-\frac{1}{(Z_{\rm TES}+R_{\rm sh}+i\omega L)I_0}\frac{Z_{\rm TES}-R_0(1+\beta)}{R_0(2+\beta)}~.
\end{eqnarray}

The TFN noise is given by
\begin{align}
\label{eq:TFNnoise}
|I_{\omega}|_{\mathrm{TFN}}^{2} =&~
2k_{B}T_c^{2}G_{\mathrm{eff}}
 \left[ \left(\frac{T_{\mathrm{bath}}}{T_c}\right)^{n+1} +1\right] \\
&\times\left[4.12-\frac{1.15}{1+\omega^{2}\tau_{1}^{2}}\right]^{ 2}\notag\\
&\times
\left|\frac{Z_{\rm TES}-R_{0}(1+\beta)}
        {(Z_{\rm TES}+R_{\mathrm{sh}}+i\omega L)(2+\beta)I_{0}R_{0}}
\right|^{2}~.\notag
\end{align}
Here, the effective thermal conductance $G_{\rm eff}$ is determined from the Joule power dissipation characterizing the power flow to the heat bath 
\begin{eqnarray}
    P_J=n^{-1}G_{\rm eff}T_c\left(1-\left(\frac{T_{\rm bath}}{T_c}\right)^n\right)~
\end{eqnarray}
and $n\simeq 5$ determined from measurements implies TES cooling by electron-phonon coupling~\cite{Hattori:2022mze}.

\begin{figure}[t]
    \centering
    \includegraphics[width=1\linewidth]{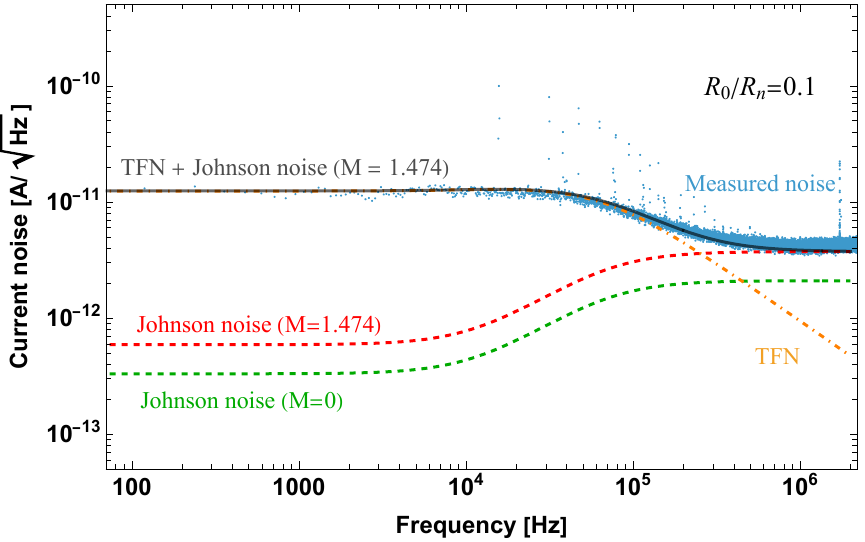}
    \caption{Current noise spectrum for TES-R considering fundamental Johnson noise (dashed) and TFN noise (dot-dashed). The total noise as measured in Ref.~\cite{Hattori:2022mze} is overlaid.}
    \label{fig:noisetot}
\end{figure}

In Fig.~\ref{fig:noisetot} we plot noise contributions together with the measured noise baseline reproduced from Ref.~\cite{Hattori:2022mze}. The total noise spectral density is given by
\begin{equation}
  |I_\omega|^2_{\rm total}=|I_{\omega}|_{\mathrm{J}}^{2}+|I_{\omega}|_{\mathrm{TFN}}^{2}~.
\end{equation}

For an ideal calorimeter the FWHM TES energy resolution can be estimated as~\cite{Irwin2005}
\begin{equation}
\label{eq:EFWHM-th}
    \Delta E_{\rm FWHM} = 2\sqrt{2\ln2}\sqrt{4k_BT_{\rm c}^2\frac{C}{\alpha}\sqrt{\dfrac{n(1+\beta)(1+M^2)}{2}}}~,
\end{equation}
where $C$ denotes the heat capacity. 
The energy resolution can be improved by lowering $T_c$, the heat capacity $C$, and the current sensitivity parameter $\beta$, or by increasing the temperature sensitivity parameter $\alpha$. Since $\alpha$ and $\beta$ are largely determined by the material properties and are challenging to modify, in this work we investigate improvement of the energy resolution by lowering $T_c$. For a fixed volume, the electronic heat capacity scales linearly with $T_c$. Hence, Eq.~\eqref{eq:EFWHM-th} implies the scaling relation $\Delta E_{\rm FWHM} \propto T_c^{3/2}$. 
Using the reference TES-R parameter values in Tab.~\ref{tableTES}, we determine $\Delta E_{\rm FWHM} \simeq 31~\mathrm{meV}$ considering $T_c = 115 ~\mathrm{mK}$ and for $M=1.474$.

In the small-signal limit, a more precise evaluation of the FWHM TES energy resolution can be obtained with the total current noise-weighted responsivity
\begin{eqnarray}
\label{eq:EFWHM-int}
    \Delta E_{\rm FWHM}=2\sqrt{2\ln 2}\left[ \int_0^\infty \frac{4|s_I(\omega)|^2}{|I_\omega|^2_{\rm total}}df \right]^{-1/2}~.
\end{eqnarray}
For the integral expression in Eq.~\eqref{eq:EFWHM-int}, the $T_c$ dependence is more complicated. From Eq.~\eqref{eq:Jnoise} and Eq.~\eqref{eq:TFNnoise} we can find the scalings $I_0 \propto T_c^{5/2}$, $G_{\rm eff} \propto T_c^{4}$, and $\tau_0,\tau_I,\tau_1 \propto T_c^{-3}$. Combining these factors, the thermal-fluctuation noise term $|I_\omega|_{\rm TFN}$ dominates the integral. Hence, the overall scaling again approximately follows $\Delta E_{\rm FWHM}\propto T_c^{3/2}$. 
With the reference parameters for TES-R in Table~\ref{tableTES}, this gives $\Delta E_{\rm FWHM}\simeq 36~ \mathrm{meV}$ $(M=1.474)$ .

In order to improve the energy resolution by approximately a factor of two requires reducing the critical temperature from $T_c = 115  ~\mathrm{mK}$ for TES-R to approximately $T_c = 72~\mathrm{mK}$ that we consider as characteristic optimized TES configuration that we call TES-O, and rescaling all $T_c$–dependent parameters in Tab.~\ref{tableTES}.
Since the thermal fall time scales as $\tau_{\mathrm{fall}}\propto T_c^{-3}$, we adopt $\tau_{\mathrm{fall}} = 16~\mu\mathrm{s}$ for the TES-O .

\begin{table}[t]
    \centering
\begin{tabular}{ r | l }
 \hline\hline
 Parameter~~ & ~~Value for TES-R (TES-O)\\[2pt] \hline
$\beta$~~                & ~~$9.22$ \\
$R_{0}$~~                & ~~$0.284~\Omega$ \\
$\mathcal{L}_{\mathrm{eff}}$~~     & ~~$53$ \\
$\tau_{1}$~~             & ~~$11.4~\mathrm{\mu s}$ ($45.6~{\rm \mu s}$)\\
$g_{2\text{-body}}$~~    & ~~$0.326$ \\
$G_{\mathrm{eff}}$~~     & ~~$6.32\times10^{-12}~\mathrm{W/K}$ ($9.95\times10^{-13}~\mathrm{W/K}$)\\
$I_{0}$~~                & ~~$6.98\times10^{-7}~\mathrm{A}$ ($2.20\times10^{-7}~\mathrm{A}$) \\
$\tau_{0}$~~             & ~~$13.9~\mathrm{~\mu s} $ ($55.6~{\rm \mu s}$)\\
$\tau_{I}$~~             & ~~$-0.27~\mathrm{~\mu s}$ ($-1.07~{\rm \mu s}$)\\
$R_{\mathrm{sh}}$~~      & ~~$1.8\times10^{-2}~\Omega$ \\
$L$~~                    & ~~$1.8\times10^{-8}~\mathrm{H}$ \\
$M^{2}$~~                & ~~$2.17$ \\
$T_c$~~     & ~~$115~\mathrm{mK}$ ($72~\mathrm{mK}$) \\
$T_{\mathrm{bath}}$~~    & ~~$7~\mathrm{mK}$ \\
$k_{B}$~~                & ~~$1.380649\times10^{-23}~\mathrm{J/K}$ \\ \hline\hline
\end{tabular}
\caption{TES design parameters used in this work considering reference TES (TES-R) as realized in Ref.~\cite{Hattori:2022mze}, and proposed optimized TES (TES-O) where energy resolution can be reduced by half.}
\label{tableTES}
\end{table}

To more precisely evaluate the TES energy resolution, we simulate the detector response and extract the resolution by applying optimal filtering to the generated data. 
Although we model the response of a single $100 ~\mu\mathrm{m}\times100~\mu\mathrm{m}$ optical TES in detail, the exposure can be increased almost linearly by tiling many pixels on a common substrate.  Modern microwave SQUID multiplexing ($\mu$MUX) architectures routinely read out $\mathcal{O}(10^{2})$ channels per feedline without significant deterioration of the intrinsic energy resolution~\cite{Nakashima:2020apl}.  To keep the discussion concrete we consider an $80$-pixel sub-array---readily achievable with a single $\mu$MUX module.  Larger arrays up to and beyond the $10^{3}$-pixel scale would simply translate the projected sensitivity curves according to the typical exposure scalings.

To highlight and isolate the fundamental noise-limited potential of the devices in our analysis for sensitivity projections we optimistically consider zero irreducible backgrounds. For sensitivity curves we quote the $95\%$ confidence level (C.L.) projection obtained from Poisson statistics, i.e. $N_{95} \simeq 3$ events. For non‑zero expected backgrounds the projection reach rescales trivially. Explicit detailed background models are beyond the scope of this work since the dominant contributions such as stray photons, radiogenic gamma and neutron fluxes, cosmogenics, etc. are highly dependent on experimental site and must be evaluated for each specific configuration.  A dedicated background mitigation strategy including shielding, underground positioning, and active vetoes hence is essential when considering particular future target experimental realizations.

\begin{figure*}[t]
    \centering
\includegraphics[width=0.46\linewidth]{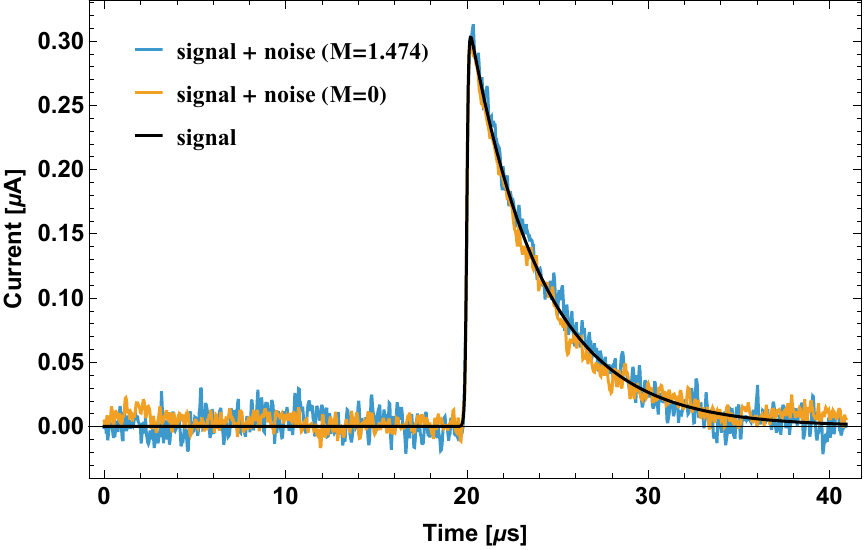}
\hspace{3em}
\includegraphics[width=0.45\linewidth]{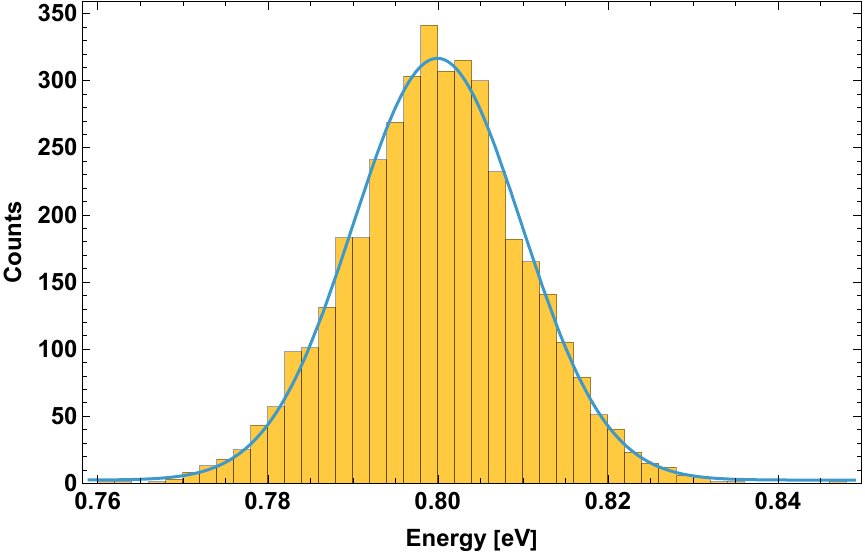}
\caption{[Left] Ideal pulse response of TES after the addition
of the electronic noise simulated with TFN and Johnson noise for reference TES parameters of Tab.~\ref{tableTES} and considering $M=1.474$ (cyan) and $M=0$ (orange) excess noise contributions.~[Right]  Calibration photon energy $E_{\rm ETF}$ fitted by optimal filtering from simulated data with TFN and Johnson noise for $M=1.474$, for 4000 realizations.}
    \label{fig:TESresponse}
\end{figure*}

\subsection{TES simulations and response fit}
\label{ssec:tessim}

To more precisely evaluate the TES energy resolution, we simulate the detector response and extract the resolution by applying optimal filtering to the generated data. 

We simulate the TES output signal as the sum of a deterministic pulse response and stochastic baseline noise. The signal component is modeled using a phenomenological template calibrated to describe single-photon events in optical TESs~\cite{Irwin2005}. The current response is expressed as
\begin{equation}
I_{\rm sig}(t)=\frac{2A_I}
       {\exp \bigl[-(t-t_{0})/\tau_{\mathrm{rise}}\bigr]
        +\exp \bigl[(t-t_{0})/\tau_{\mathrm{fall}}\bigr]},
        \label{eq:signal}
\end{equation}
where $A_I$ is half the peak amplitude, $t_0$ is the trigger time, $\tau_{\mathrm{rise}}$ is the rise time governed bythe L/R electrical time constant, and $\tau_{\mathrm{fall}}$ is the decay time determined by the ETF and the effective thermal conductance $G_{\mathrm{eff}}$ to the bath. This template accurately describes TES signals and provides a well-defined pulse kernel for matched filtering.

In Fig.~\ref{fig:TESresponse} we display an example of the analytic template evaluated for a 80 ns sampling grid using best fit parameters $( \tau_{\mathrm{rise}}=50~{\rm ns},~t_0=20~{\rm \mu s}, ~\tau_{\mathrm{fall}}=4~{\rm \mu s})$ 
and signal amplitude $A_I=1.62 \times 10^{-7}$ A for TES-R as well as $( \tau_{\mathrm{rise}}=50~{\rm ns},~t_0=20~{\rm \mu s}, ~\tau_{\mathrm{fall}}=16~{\rm \mu s})$
and signal amplitude $A_I=5.54\times 10^{-8}$ A for TES-O calculated from
\begin{eqnarray}
    E_{\rm ETF}=-\int I_0R_0 I_{\rm sig}(t)dt
\end{eqnarray}
with the calibration photon energy $E_{\rm ETF}=0.8~{\rm eV}$ used in Ref.~\cite{Hattori:2022mze}. We note that current can be readily converted to voltage, considering a factor of 270 k$\Omega$.

To simulate the baseline noise, we construct the time-domain noise signal from the theoretical current noise spectral density $|I_\omega|_{\mathrm{total}}^2$, which includes both the fundamental Johnson noise and TFN, as discussed above. For each Fourier frequency bin $f_i$ up to the Nyquist frequency $f_{\rm Ny} = 12.5  ~\mathrm{MHz}$, we generate two independent Gaussian random variables with zero mean and unit variance, and scale them by $(|I_\omega(f_i)|^2 \Delta f/2)^{1/2}$ to define the real and imaginary parts of the Fourier-domain noise. The negative-frequency components are set by complex  conjugation to ensure a real time-domain signal. An inverse fast Fourier transform (FFT) yields the time-domain noise stream $\delta I(t)$.
Adding the noise to the pulse template gives
\begin{equation}
I(t)=I_{\mathrm{sig}}(t)+\delta I(t)~.
\end{equation} 

To extract the energy resolution, we apply an optimal filter in the frequency domain~\cite{McCammon:2005im}. The optimal filter is designed to minimize the variance of the amplitude estimate for a known signal template in the presence of stationary Gaussian noise. Denoting the discrete Fourier coefficients of the signal waveform as $\tilde I_k$, the unit-normalized Fourier template as $\tilde h_k = \tilde V_{{\rm sig},k}/A$, and the noise spectral density as $S_{I,k}$, the maximum-likelihood estimator for the pulse amplitude is
\begin{equation}
\hat A=\frac{\displaystyle\sum_{k}\tilde I_k\tilde h_k^{*}/S_{I,k}}
             {\displaystyle\sum_{k}|\tilde h_k|^{2}/S_{I,k}},
\end{equation}
and its variance   
\begin{equation}
\sigma_A^{2}=\dfrac{1}{\sum_{k}|\tilde h_k|^{2}/S_{I,k}}.
\end{equation}
This estimator yields both the reconstructed pulse amplitude and the minimum statistical uncertainty for a linear measurement. We can then convert the amplitude variance to an energy resolution.

\begin{table}[]
    \centering
    \begin{tabular}{lrr}
     \hline\hline
Excess Johnson Noise (M) & 0 & \quad 1.474\\
\hline
TES-R Eq.~\eqref{eq:EFWHM-th} ($\Delta E$, eV) & \quad\quad 0.023 & 0.031 \\  
TES-R Eq.~\eqref{eq:EFWHM-int} ($\Delta E$, eV) & \quad\quad 0.027 & 0.036 \\  
TES-R Optimal Filtering ($\Delta E$, eV) & \quad\quad 0.026 & 0.031 \\ 
\hline
TES-O Eq.~\eqref{eq:EFWHM-th} ($\Delta E$, eV) & \quad\quad 0.011 & 0.015 \\  
TES-O Eq.~\eqref{eq:EFWHM-int} ($\Delta E$, eV) & \quad\quad 0.013 & 0.018 \\  
TES-O Optimal Filtering ($\Delta E$, eV) & \quad\quad 0.012 & 0.016 \\ 
     \hline\hline
    \end{tabular}
    \caption{Comparison of energy resolution $\Delta E_{\mathrm{FWHM}}$ obtained using Eq.~\eqref{eq:EFWHM-th}, the small-signal expression Eq.~\eqref{eq:EFWHM-int},  and full optimal filtering simulation. TES-R denotes reference TES at $T_c = 115~\mathrm{mK}$, while TES-O refers to a lower-$T_c$ design at $72~ \mathrm{mK}$.}
    \label{tab:EFWHM}
\end{table}

We simulate 4000 pulse response realizations and apply optimal filtering procedure with realistic noise conditions for TES-R and TES-O. In Tab.~\ref{tab:EFWHM}, we compare the results of optimal filtering against the analytic small-signal limit from Eq.~\eqref{eq:EFWHM-int} for both TES-R with $T_c = 115~ \mathrm{mK}$  and TES-O with $T_c = 72~  \mathrm{mK}$. We include two cases for the excess Johnson noise factor $M = 0$ and $M = 1.474$, as inferred from experimental fits.

The agreement between the analytic and numerical results confirms the robustness of the optimal filter approach. Furthermore, our simulations reproduce the experimentally measured single-photon resolution of $\Delta E_{\mathrm{FWHM}} \simeq 67\ \mathrm{meV}$ from Ref.~\cite{Hattori:2022mze}, validating the TES design and confirming that its performance is fundamentally limited by Johnson noise and TFN under low-background conditions, such as those expected in underground laboratories.

For sensitivity projections in DM searches, we model the TES as a near-ideal calorimeter. The interaction rates are calculated in terms of true deposited energy $E_{\mathrm{dep}}$ and convolved with the detector response to obtain the measured energy distribution. From our simulations, the baseline variance $\sigma_A^2$ for zero-energy events is determined using the same optimal filter method. Assuming Gaussian statistics, the single-event energy resolution is $\Delta E_{\mathrm{FWHM}} = 2.355\sigma_A$. We adopt a $3\sigma$ threshold criterion for triggering, giving a conservative but realistic detection threshold of $E_{\mathrm{th}} = 3\sigma_A \simeq 0.1  \mathrm{eV}$, which we use throughout our sensitivity estimates.

 \begin{figure*}[t]
    \centering
    \includegraphics[width=0.475\linewidth]{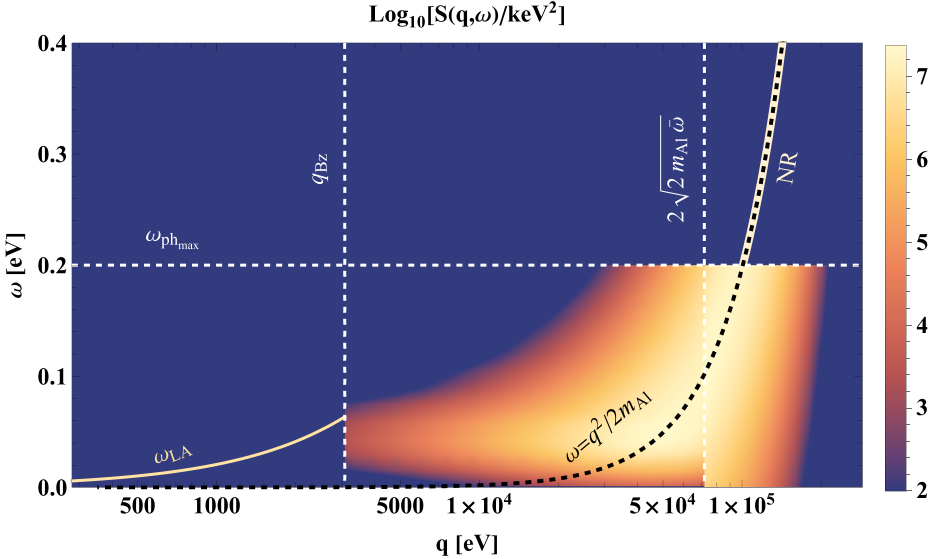} \hspace{2em}
    \includegraphics[width=0.475\linewidth]{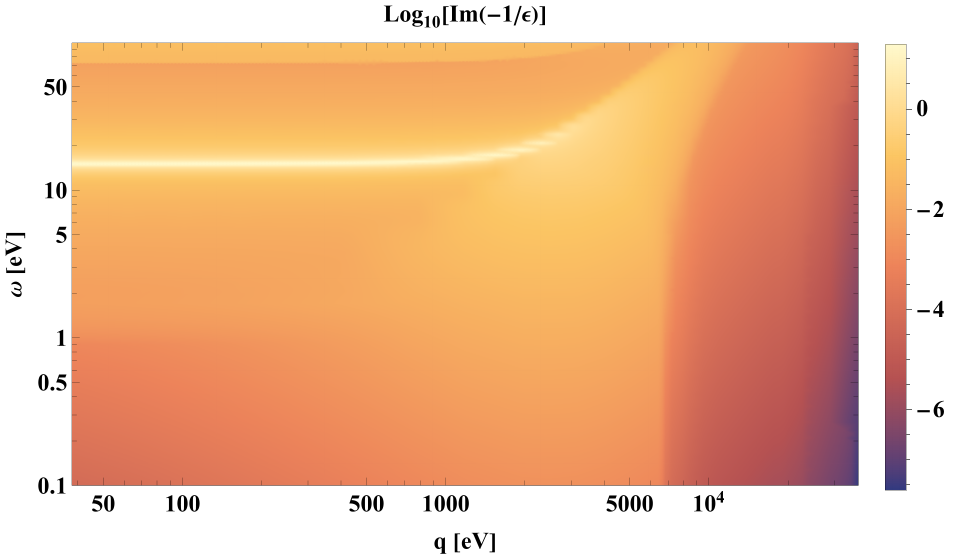}
    \caption{[Left] Density plot of the nuclear dynamic structure factor $S_n(q,\omega)$ for aluminum. [Right] Density plot of the energy loss function ${\rm Im}[-1/\epsilon(\omega,q)]$ for aluminum.}
    \label{fig:alepsilon}
\end{figure*}
 
\section{TES Target Properties}
\label{subsec:AlResponse}

To explore DM interactions, we consider TES itself as Al target, and also consider additional Al absorbers. 
The interactions of DM with a condensed-matter target are governed by two complementary response functions. The dynamic structure factor $S(q,\omega)$ encodes microscopic density–density correlations, and controls processes considering energy $\omega$ and momentum $q$ are transferred to the medium. On the other hand,  the energy–loss function (ELF) ${\rm Im}\bigl[-1/\epsilon(q,\omega)\bigr]$  is a macroscopic quantity derived from the complex longitudinal dielectric function $\epsilon$
and governs the dissipation of an external electromagnetic field. Both quantities can be expressed in terms of equilibrium correlators of number-density operators. They depend only on the intrinsic properties of the material and not on the specific probe that excites it, such as DM interactions.

In the remainder of this section we summarize the Al response needed for our DM-scattering and DM-absorption calculations—namely $S_n(q,\omega)$ for nuclei, $S_e(q,\omega)$ for electrons and dielectric function $\epsilon(q,\omega)$. For numerical computations we employ~\texttt{DarkELF}~\cite{Knapen:2021bwg} code, which combines experimental data with ab-initio calculations.

\subsection{Dynamic structure factor}
\label{ssec:dynamics}

For a target material of volume $V_T$ the dynamic structure factor for species $i$, with $i=e,n$ for electrons and nuclei, respectively, is defined by (see App.~\ref{app:calc} for details)
\begin{align} \label{eq:Si_omega}
  S_i(q,\omega)
  &=
  \frac{1}{V_T}
  \int_{-\infty}^{\infty}dt 
  e^{i\omega t}~
  \bigl\langle n_i(q,t)~n_i(-q,0) \bigr\rangle \\
  &=\frac{2\pi}{V_T}\sum_{m,n}
    p_m \left|\left<m|n_i(q)|n\right>\right|^2\delta(\omega+E_m-E_n), \notag 
\end{align}
where $n_i(q,t)$ is the Fourier transform of the number density $n_i(x,t)$, $q = |\textbf{q}|$ is the momentum transfer and $\omega$ is the energy transfer.
The states $|m\rangle$ are many-body eigenstates of the unperturbed Hamiltonian with energies $E_m$, and $p_m=e^{-E_m/T}/\sum_n e^{-E_n/T}$ are the Boltzmann occupation probabilities at temperature $T$. Eq.~\eqref{eq:Si_omega} is valid for electrons and for the center-of-mass motion of the atomic cores. The latter can include single-phonon and multi-phonon excitations, as well as recoil of the individual nuclei in the impulse limit. For our analysis we consider scenarios where DM is dark photon or DM couples to material through a scalar/vector mediator. In these cases, the dynamic structure factor of the electron/nucleus number density captures the effect of DM absorption/scattering in the non-relativistic limit.

If the DM or its mediator couples to the electronic charge density, the relevant quantity is electron dynamic structure factor $S_e(q,\omega)$. The fluctuation–dissipation theorem connects this microscopic quantity to the macroscopic longitudinal dielectric function $\epsilon(q,\omega)$ through~\cite{PhysRev.113.1254}
\begin{equation} \label{eq:Sqw_eps}
  S_{e}(q,\omega)
  =
  \frac{q^{2}}{2\pi\alpha_e}
  \frac{{\rm Im}\bigl[-1/\epsilon(q,\omega)\bigr]}
       {1-\exp(-\omega/T)}~,
\end{equation}
where $\alpha_e =e^2/(4\pi)$ is the fine structure constant and
${\rm Im}\bigl[-1/\epsilon(q,\omega)\bigr]$ is ELF\footnote{
We note that the measured ELF in experiments can contain contributions from phononic absorption.  For $\omega$ higher than the typical phonon energy, interactions with electrons give dominant contribution. 
}.
The ELF quantifies dissipation under an external driving field and is therefore appropriate quantity to employ when Standard Model (SM) photon, dark photons or  millicharged DM particles interact with the target.  

The factor $1/\epsilon$ in Eq.~\eqref{eq:Sqw_eps}, rather than $\epsilon$ itself, can be understood from standard electrodynamics of interfaces~\cite{Dressel_Grüner_2002}.
Across the boundary between the material and vacuum the displacement field
$D_i=\epsilon_{ij}E_j$ is continuous, whereas the microscopic electric field is not.
Writing the fields just inside (``in'') and outside (``out'') the sample one has
$\epsilon E^{\mathrm{(in)}}_i  = D^{\mathrm{(in)}}_i 
                              =  D^{\mathrm{(out)}}_i 
                              =  E^{\mathrm{(out)}}_i .
$ 
Hence, the external driving field is $E^{\mathrm{(out)}} = E^{\mathrm{(in)}}/\epsilon$.
The electromagnetic energy stored per unit volume inside the medium is
$
u_{\mathrm{in}}
  = \dfrac{\epsilon}{2} \bigl|E^{\mathrm{(in)}}\bigr|^{2}
  =            \dfrac{1}{2\epsilon} \bigl|E^{\mathrm{(out)}}\bigr|^{2}.
$
Dissipation therefore scales with $\mathrm{Im} [1/\epsilon]$, which is the quantity that appropriately describes energy loss in response to an external  field.
A systematic derivation of this result in the context of DM absorption is given in Ref.~\cite{Mitridate:2021ctr}.

If DM or the mediator couples instead to nucleons, the relevant quantity is the nuclear structure factor $S_n(q,\omega)$.
With the number density of nucleons $n$ being $n_n({\bf x}) = \sum_\ell A \delta({\bf x - x_\ell})$, where $A = 27$ is the mass number of nucleus for Al and $\ell$ labels the lattice sites, Eq.~\eqref{eq:Si_omega} becomes\footnote{
    Here we assume that a unit cell contains one atom, which is true for Al. Consequently, there are no optical–phonon branches.
}
\begin{align}
    S_n(\omega, q) = \frac{2\pi A^2}{V_T}\sum_{f}\left|\sum_{\ell} \left<f\right| e^{i{\bf q\cdot x_\ell}} \left|0\right> \right|^2 \delta(\omega-E_f),
\end{align}
where $|0\rangle$ is the $T \to 0$ ground state and $V_{T}$ is the total target volume. For light DM the intermediate state $|f\rangle$ can generally contain one or more phonons. Accordingly, $S_{n}$ must be evaluated in various kinematic regimes~\cite{Trickle:2019nya,Campbell-Deem:2022fqm,Griffin:2024cew}. 

For low momentum transfers wihtin the first Brillouin zone, $q\lesssim q_{\mathrm{BZ}}= 2\pi/a_{\mathrm{lat}}$ with $a_{\mathrm{lat}}\simeq0.44~\mathrm{nm}$ for Al, the response is dominated by creation of a single longitudinal‑acoustic (LA) phonon,
\begin{eqnarray}
    S_{n}^{\rm (LA)}(q,\omega)&=&\frac{2\pi A^2}{V_{\rm cell}}\frac{q^2}{2m_{\rm Al}\omega^{\rm (LA)}_q}\delta(\omega-\omega^{\rm (LA)}_q)
\end{eqnarray}
where $V_{\mathrm{cell}}$ is the primitive‑cell volume, $m_{\mathrm{Al}}=25.0~\mathrm{GeV}$ the Al atomic mass, and $\omega_{q}^{\text{(LA)}} = q c_{\text{LA}} $ represents the dispersion relation of the LA mode with sound speed $c_{\text{LA}}\simeq 6.2~\mathrm{km~s^{-1}}$.

For momentum transfers exceeding the first Brillouin zone boundary,
$q>q_{\rm Bz}$, single–phonon creation is no longer
kinematically favored and the response is instead governed by
multiphonon and, at still higher $q$, by nuclear recoil
processes. For $q_{\rm Bz}<q<2\sqrt{2m_{\rm Al} \bar\omega}$ the scattering excites
$n\ge1$ phonons. In the incoherent approximation
\begin{align}
\label{eq:Smulti}
    S^{\rm (multi)}_n(q,\omega) =&~\frac{2\pi A^2}{V_{\rm cell}}e^{-2W(q)}\sum_n\left( \frac{q^2}{2m_{\rm Al}}\right)^n\frac{1}{n!}\\
    &\times  \left(\prod_{j}\int\frac{d\omega_jD(\omega_j)}{\omega_j}\right)\delta\left(\sum_j \omega_j-\omega\right) ~.\notag
\end{align}
Here, $D(\omega)$ denotes the phonon density of states. The Debye–Waller factor that resums virtual phonon exchange $W(q)$ is given by
\begin{eqnarray}
    W(q)=\frac{q^2}{4m_{\rm Al}}\int\frac{d\omega' D(\omega')}{\omega'}~.
\end{eqnarray}
Expression Eq.~\eqref{eq:Smulti} is valid for momentum transfers in the range $q_{\rm Bz}<q<2\sqrt{2m_{\rm Al} \bar{\omega}}$, where $\bar{\omega}$ is a representative phonon energy - for Al we take the average acoustic phonon energy $\bar{\omega}=\int d\omega~ D(\omega)\omega=25$~meV~\cite{Knapen:2021bwg}.

\begin{figure*}[t]
    \centering
    \includegraphics[width=0.475\linewidth]{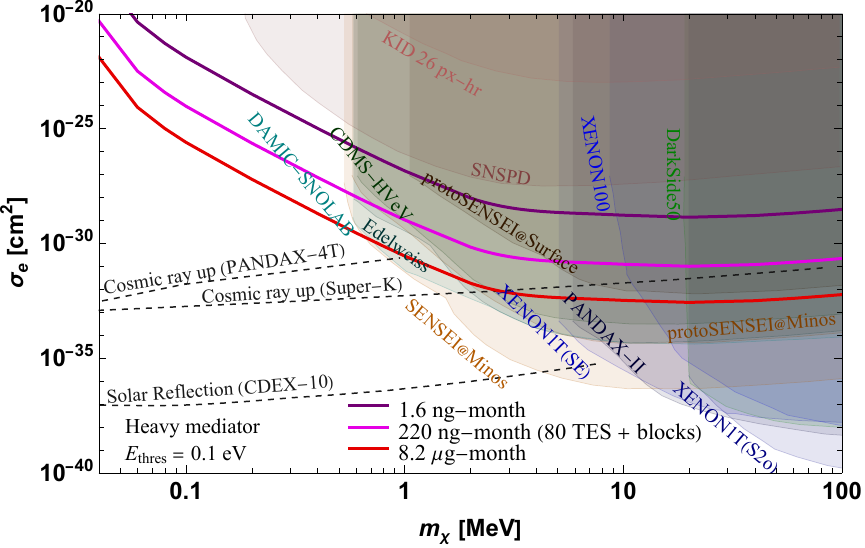} \hspace{1em}
    \includegraphics[width=0.475\linewidth]{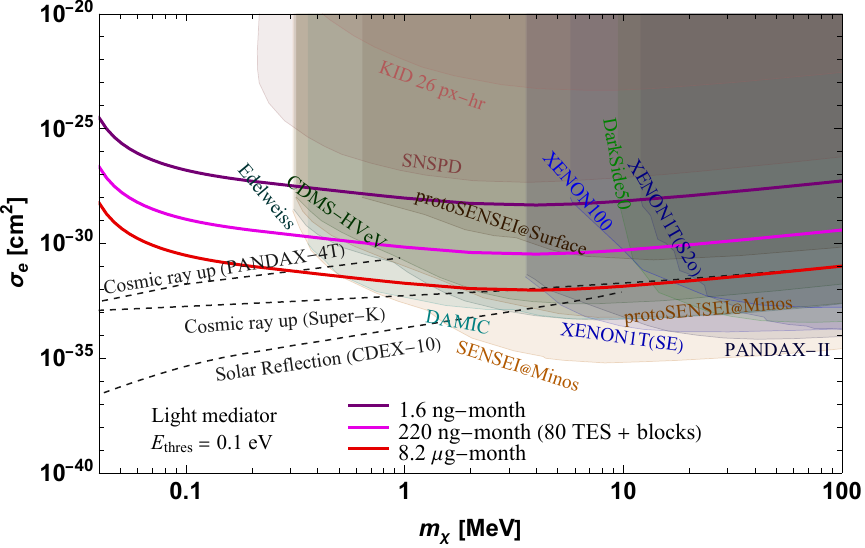}
    \caption{[Left] Sensitivity projections for TES detection of DM-electron scattering considering heavy mediator at the $95~\%$ C.L., for various exposures (solid lines). The sensitivity is calculated using ELF calculated from the Mermin method, with energy threshold $E_{\rm th}=0.1~{\rm eV}$. 
    The shaded regions indicate constraints derived from KID~\cite{Gao:2024irf} and SNSPD~\cite{Hochberg:2021yud} data, from XENON100~\cite{XENON:2016jmt}, XENON1T (single-electron (SE) and S2-only (S2o))~\cite{PhysRevLett.123.251801,PhysRevD.106.022001}, PANDAX-II~\cite{PhysRevLett.126.211803}, DarkSide-50~\cite{PhysRevLett.121.111303}, EDELWEISS~\cite{PhysRevLett.125.141301}, DAMIC at SNOLAB~\cite{PhysRevLett.123.181802}, CDMS-HVeV~\cite{PhysRevLett.121.051301,PhysRevD.102.091101}, SENSEI~\cite{PhysRevLett.121.061803,PhysRevLett.122.161801,PhysRevLett.125.171802}. [Right] DM-electron scattering with light mediator.}
    \label{fig:DMelectron}
\end{figure*}

At still higher momentum, $q>2\sqrt{2m_{\rm Al}\bar{\omega}}$, energy and momentum are deposited abruptly and lattice ions behave as quasi-free particles. In this impulse approximation the response is a Gaussian centered on the free recoil energy 
\begin{align}
    S_n^{\rm (IA)}(q,\omega)=&~\frac{2\pi A^2}{V_{\rm cell}}\frac{1}{\sqrt{2\pi}\Delta(q)}\\
    & \times\exp\left[{-\frac{(\omega-q^2/2m_{\rm Al})^2}{2\Delta^2(q)}}\right]\notag
\end{align}
with variance $\Delta^{2}(q)=q^{2}\bar{\omega}/(2m_{\rm Al})$.
  
For energy–momentum $(q, \omega)$ transfers that probe the internal structure of the nucleus, the scattering reduces to incoherent nuclear recoil modified by the nuclear form factor,
\begin{equation}
    S^{\rm (NR)}(q,\omega)=\dfrac{2\pi A^2}{V_{\rm cell}}\frac{3j_1(qr_{\rm Al})}{(qr_{\rm Al})}e^{-(qs)^2}\delta(\omega-\dfrac{q^2}{2m_{\rm Al}})~,
\end{equation}
where $j_{1}$ is the spherical Bessel function, $r_{\rm Al}\simeq1.14 A^{1/3}~\text{fm}\simeq 3.42$~fm is the effective nuclear radius and $s\simeq0.9$~fm parametrizes the surface-diffuseness ``skin-depth'' of the nucleus.

In Fig.~\ref{fig:alepsilon} we display
$S_{n}(q,\omega)$ for several representative $q$-values.  
The low-$\omega$ acoustic ridge dominates DM scattering for $m_{\chi}\lesssim50$~MeV, whereas multi-phonon and impulse-approximation continua control the response at higher masses.

\subsection{Dielectric function}

Al is a simple metal with no electronic band gap. The ELF in the electronic regime, $\operatorname{Im}\bigl[-1/\epsilon(q,\omega)\bigr]$, can be deduced at small momentum from reflection–electron–energy–loss and optical data by fitting to Mermin oscillator model~\cite{Mermin:1970zz}. Those data provide the complete complex dielectric function in the $q \to 0$ limit for Eq.~\eqref{eq:Sqw_eps}. For larger momenta, where direct measurements are more sparse, we match the empirical fit onto a Lindhard model calculation that reproduces the electron–hole continuum.

The fluctuation–dissipation theorem makes the imaginary part of the dielectric function the central quantity for DM–electron interactions, because it is directly related to the electron dynamic structure factor as through Eq.~\eqref{eq:Sqw_eps}. The dielectric response of the conduction electrons, which directly encodes the medium’s optical and energy dissipation properties, characterizes Al sensitivity to DM energy deposits.

The zero-temperature Lindhard dielectric function for a free electron gas with Fermi velocity $v_F$~(e.g.~\cite{Dressel_Grüner_2002,Fetter:2003,Quinn:2018}) is given by
\begin{equation}
    \epsilon_{\rm L} (q, \omega) =1+\dfrac{3\omega^2_p}{q^2v_F^2}\lim_{\eta\rightarrow 0}\left[f\left(\frac{\omega+i\eta}{q v_F},\dfrac{q}{2m_ev_F}\right)\right]
\end{equation}
where
\begin{eqnarray}
    &f(u,z)=\dfrac{1}{2}+\dfrac{1}{8z}[g(z+u)+g(z-u)]\\
    &g(x)=(1-x^2)\log\left[\dfrac{1+x}{1-x}\right] \notag
\end{eqnarray}
and
$\omega_{p}=15.3~\text{eV}$ is the Al bulk plasma frequency  and $m_{e}$ is the electron mass.

Dissipation from impurities and electron–phonon scattering can be incorporated through the Mermin dielectric function prescription~\cite{Mermin:1970zz}, which generalizes the Lindhard model by including a finite electron collision rate $\gamma$ 
\begin{equation} \label{eq:Mermin}
  \epsilon(q,\omega)
  =
  1
  +
  \frac{\bigl[1+i\gamma/\omega\bigr]
        \bigl[\epsilon_{\mathrm{L}}(q,\omega+i\gamma)-1\bigr]}
       {1+\dfrac{i\gamma}{\omega}
        \dfrac{\epsilon_{\mathrm{L}}(q,\omega+i\gamma)-1}
             {\epsilon_{\mathrm{L}}(q,0)-1}},
\end{equation}
where $\gamma$ is the phenomenological electron–collision rate. 
 
In Fig.~\ref{fig:alepsilon} we display the resulting nuclear structure factor $S_n(q,\omega)$ calculated as described in Sec.~\ref{ssec:dynamics} and $\operatorname{Im}\bigl[-1/\epsilon(q,\omega)\bigr]$ calculated considering Mermin dielectric function for several representative momenta. A pronounced plasmon peak appears near $\omega\simeq\omega_{p}$, which also dominates dark photon absorption. The broad Landau tail below $1~\text{eV}$ is governed by sub-eV electron recoils as relevant for light DM.
  
\section{Dark Matter Searches} 
\label{sec:dme}

Having developed optical TES characterization, we now turn to exploring DM detection strategies that leverage its sensitivity. 

\subsection{Electron scattering}

We consider DM interactions with electrons through a mediator particle in our optical Al TES target.  As discussed above, the relevant in-medium scattering rate is governed by the ELF of the material, $\mathrm{Im}[-1/\epsilon(\omega, q)]$. 

One motivated scenario is DM-electron scattering through a vector (dark photon) mediator $A_{\mu}'$. 
The Lagrangian is
\begin{align}
\mathcal{L} \supset
& -\frac{1}{4} F^{\mu\nu}F_{\mu\nu}
- \frac{1}{4} F'^{\mu\nu}F'_{\mu\nu}
+ i \bar\chi \gamma^\mu \partial_\mu\chi + m_\chi\bar\chi\chi \nonumber \\
&+ \frac{1}{2} m_{V}^2 A'_{\mu}A'^{\mu}
 + \left(g_e J_e^{\mu} + g_{\chi} J^{\mu}_{\chi}\right) A'_{\mu},
 \label{eq:L_scat_e}
\end{align}
where $F_{\mu\nu}$ and $F'_{\mu\nu}$ are the field strengths of the photon and dark photon, $\chi$ is DM fermion,
$m_{V}$ the dark photon mass and $m_{\chi}$ is the DM fermion mass, $J_e^\mu$ the electron current and $J_\chi^\mu$ the DM current with coupling constants $g_e$ and $g_\chi$.  In-medium photon-$A'$ mixing modifies the propagator through the longitudinal dielectric function $\epsilon(\omega, q)$ of the target. Here, these effects are taken into account by the ELF.
Although we focus on the vector mediator below, analogous calculation also applies to the case of scalar mediator, after reinterpreting $m_V$ as the scalar mass $m_\phi$ and $g_\chi$ $(g_e)$ as the scalar-DM (scalar-electron) couplings.
See Ref.~\cite{Knapen:2017xzo} for a review of such light DM coupled through mediator.

For a non-relativistic DM particle scattering with incoming velocity $\mathbf{v}$ and outgoing velocity ${\bf v}'$, 
the energy transfer is given by $\omega = \frac{1}{2}m_\chi(v^2- {v'}^{2})$ and the momentum transfer by ${\bf q} = m_\chi({\bf v} - {\bf v}')$, hence there is a relation
\begin{equation} \label{eq:DeltaE}
  \omega
  =
  \mathbf{q}\cdot\mathbf{v}
  -\frac{q^{2}}{2m_{\chi}}~.
\end{equation}
This implies the kinematic threshold  
\begin{equation}
v_{\rm min}(\omega, q) = \frac{\omega}{q} + \frac{q}{2 m_\chi}.
\end{equation}

For the DM halo velocity distribution $f(\mathbf{v})$, 
we use the truncated Maxwell-Boltzmann distribution following the Standard Halo Model~\cite{Drukier:1986tm}
\begin{eqnarray}
    f(\mathbf{v})=\frac{1}{N_0}e^{-((\mathbf{v}-\mathbf{v}_e)/v_0^2)}\Theta (v_{\rm esc}-|\mathbf{v}+\mathbf{v}_e|)~,
\end{eqnarray}
where $N_0$ is the normalization factor, $|\mathbf{v}_e|=240~{\rm km/s}$ is the velocity of the Earth in the Galactic rest frame,  $v_0=230~{\rm km/s}$ is the average speed of DM in the halo, and $v_{\rm esc}=600~{\rm km/s}$ is the Galactic escape velocity at the Sun's location. One can define the inverse mean velocity weighted by the halo distribution
\begin{equation} \label{eq:eta}
  \eta(v_{\min})
  =
  \int_{v>v_{\min}} \mathrm{d}^{3}v
  ~ \frac{f(\mathbf{v})}{v}.
\end{equation}

\begin{figure}[t]
    \centering
    \includegraphics[width=1\linewidth]{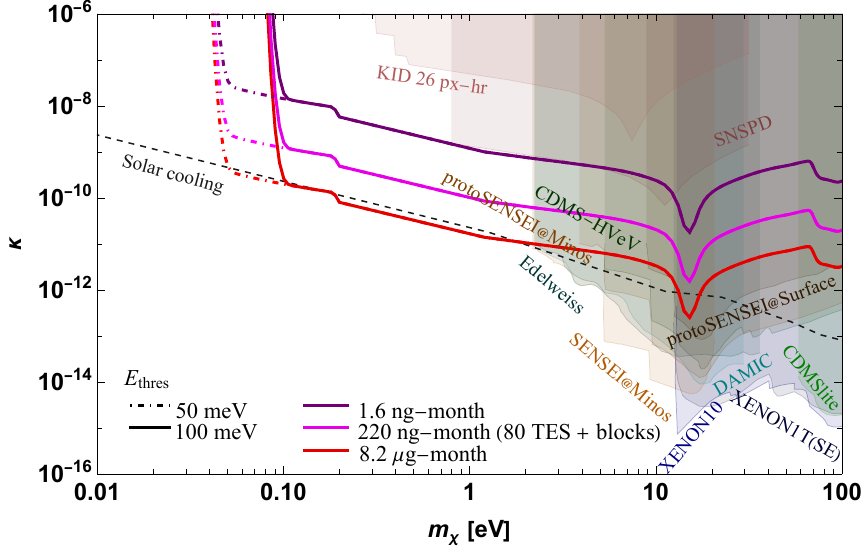}
    \caption{Sensitivity projections for TES detection of DM absorption, for different exposures and energy thresholds (dot-dashed and solid lines). The shaded regions indicate constraints derived from KID~\cite{Gao:2024irf} and SNSPD~\cite{Hochberg:2021yud} data, from XENON1T~\cite{PhysRevLett.123.251801,PhysRevD.106.022001}, EDELWEISS~\cite{PhysRevLett.125.141301}, DAMIC at SNOLAB~\cite{PhysRevLett.118.141803},  SENSEI~\cite{PhysRevLett.121.061803,PhysRevLett.122.161801,PhysRevLett.125.171802}, XENON10, and CDMSlite~\cite{Bloch:2016sjj}.}
    \label{fig:DMabsorption}
\end{figure}

The DM scattering rate per unit time per detector mass (e.g. in units of ${\rm s^{-1}~kg^{-1}}$) is given by (see App.~\ref{app:calc} for detail) 
\begin{align}
\label{eq:R_scatter}
	R &= \frac{\rho_\chi}{\rho_T m_\chi} \int d^3v f(\mathbf{v}) \int \frac{d^3q}{(2\pi)^3}
	\left(\frac{g_e g_\chi}{q^2 + m_V^2}\right)^2 S_e(\omega, \mathbf{q}) \nonumber\\
	&=\frac{\rho_\chi}{\rho_T m_\chi} \int d^3v f(\mathbf{v}) \int \frac{q^2 d^3q}{(2\pi)^3}
	\frac{2g_e^2 g^2_\chi}{e^2(q^2 + m_V^2)^2} {\rm Im}\left[\frac{-1}{\epsilon(\omega,q)}\right]
\end{align}
considering isotropic targets. Here, $\rho_\chi = 0.4~\mathrm{GeV~cm^{-3}}$ is taken as the local DM density~(e.g.~\cite{Staudt_2024}) and $\rho_T$ is the mass density of the target material, and we have taken the zero-temperature limit.
We note that $\omega$ in Eq.~\eqref{eq:R_scatter} should be understood as a function of ${\bf q}$ and ${\bf v}$ through Eq.~\eqref{eq:DeltaE}.
Eq.~\eqref{eq:R_scatter} can be also conveniently stated in terms of the reference DM-electron cross section as~\cite{Essig:2011nj,Essig:2015cda,Knapen:2021run}
\begin{align}
	R&= \frac{\rho_\chi}{\rho_T m_\chi} \frac{\bar\sigma_e}{2\alpha_e\mu_{\chi e}^2} \int d^3v f(v)
	\nonumber\\
	&~~~\times \int \frac{d^3q}{(2\pi)^3} q^2 \left|F_{\rm med}(q)\right|^2   {\rm Im}\left[\frac{-1}{\epsilon(\omega,q)}\right]~,
    \label{eq:Rscat}
\end{align}
where the mediator form factor is 
\begin{equation}
\label{eq:Fmed}
F_{\rm med}(q)=\frac{m_{V}^{2}+q_{0,e}^2}{m_{V}^{2}+q^{2}}~,
\end{equation} 
and the DM-electron reference cross-section is
\begin{equation} \label{eq:sigma_e_bar}
  \bar{\sigma}_{e}
  =
  \frac{g_e^2 g_\chi^2 \mu_{\chi e}^{2}}
{\pi\bigl(m_{V}^{2}+q_{0,e}^2\bigr)^{2}}~,
\end{equation}
with reference momentum $q_{0,e} = \alpha_e m_e$ and $\mu_{\chi e}$ being the reduced mass of the electron and DM.

\begin{figure}[t]
    \centering
    \includegraphics[width=1\linewidth]{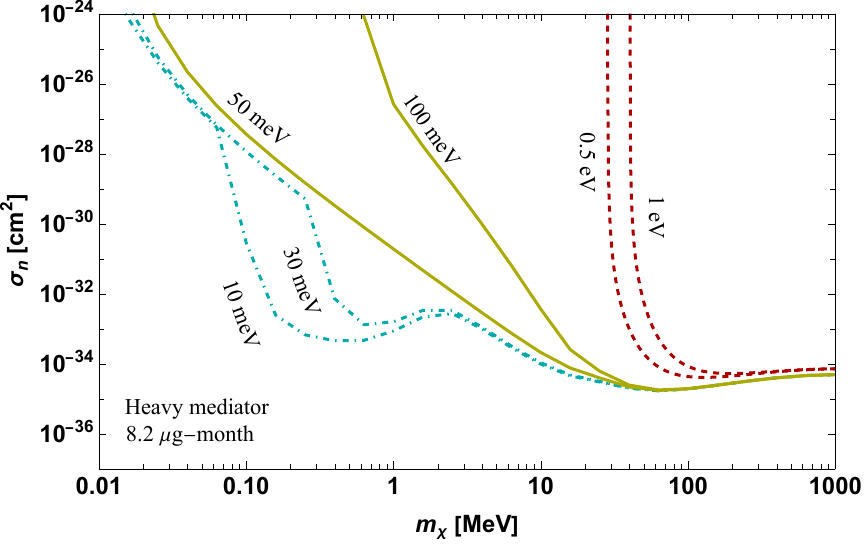}
    \caption{DM-nucleon coupling with phonon regimes, for $8.2~\mu$g-month exposure, considering different thresholds.}
    \label{fig:DMphonon}
\end{figure}

To estimate the DM interaction event rate, we need to account for the detector's energy threshold $E_{\rm th}$. This is implicitly encoded in Eq.~\eqref{eq:Rscat} through the integration limits over ${\bf q}$ and ${\bf v}$. For clarity, it is convenient to rewrite the expression as follows. We insert a trivial identity,
$
1 = \int d\omega\, \delta\left(\omega - {\bf q \cdot v} +  q^2/2m_\chi\right),
$
and exchange the order of integration. Introducing the angle $\theta$ between ${\bf q}$ and ${\bf v}$, we perform the $\theta$ integral using the delta function, which yields a Jacobian factor $(qv)^{-1}$. After determining the physical integration ranges of $\omega$ and $q$, we arrive at the expression presented in Ref.~\cite{Campbell-Deem:2022fqm} 
\begin{align}
\label{eq:dRdo_to_R}
    R = \int_{E_{\rm th}}^{\omega_{\rm max}} d\omega \frac{dR}{d\omega}~,
\end{align}
where
\begin{align}
\label{eq:dRdomega_e}
    \frac{dR}{d\omega} &= \frac{\rho_\chi}{\rho_T m_\chi} \frac{\bar\sigma_e}{2\alpha_e\mu_{\chi e}^2}
    \int_{q_-}^{q_+} \frac{q^3dq}{(2\pi)^2} \left|F_{\rm med}(q)\right|^2 \nonumber\\
    &~~~\times \mathrm{Im}\left[\frac{-1}{\epsilon(\omega, q)}\right]
\eta\left(v_{\rm min}(\omega, q)\right)~.
\end{align}
Here $\eta\left(v_{\rm min}\right)$ is given by Eq.~\eqref{eq:eta}, $\omega_{\rm max} = m_\chi v_{\rm max}^2/2$ with $v_{\rm max}$ being the maximum DM velocity on Earth and
\begin{align}
    q_{\pm}(\omega) = m_\chi v_{\rm max} \left( 1\pm \sqrt{1-\frac{2\omega}{m_\chi}}\right).
\end{align}

Our high-resolution TES is capable of resolving energy depositions as low as $E_{\rm th} = 0.1~\mathrm{eV}$. The measured ELF of aluminum exhibits a Landau damping continuum below $\sim 10$~eV. This enables sensitivity to DM scattering-induced sub-eV electron excitations. The Al plasmon peak near $\omega_p \sim 15~\mathrm{eV}$ is kinematically inaccessible for MeV-scale DM scattering, but can be relevant for DM absorption processes discussed in the next subsection.

In Fig.~\ref{fig:DMelectron}, we present projected sensitivities for benchmark exposures and both heavy and light mediator scenarios of DM interactions, corresponding respectively to flat and Coulomb-like $F_{\rm med}(q)$ form factors. 
The smallest considered exposure, 1.6 ng-month, corresponds to a single optical TES pixel and a month livetime of experiment setup operation. The intermediate benchmark of ``80 TES + blocks'' corresponds to an 80-pixel sub-array, augmented by Al collection blocks attached to each TES to increase the target mass without significantly compromising energy resolution. These blocks are assumed to contribute an additional factor of $\sim 10$ in mass per pixel, consistent with designs that enhance absorber volume while maintaining thermal coupling. The total effective target mass for this configuration is approximately 220~ng. Finally, the $8.2~\mu$g-month exposure represents a forward-looking scenario achievable with large multiplexed TES arrays (e.g. $\mathcal{O}(10^3)$ pixels), extended integration time, and optimized absorber engineering. These values are chosen to span current capabilities and optimistic but realistic near-future projections for quantum calorimetric rare-event searches.

The excellent energy resolution and low threshold of the TES setup enable sensitivity to previously unexplored parameter space for $m_\chi \sim 0.1$--$10~ \mathrm{MeV}$, reaching cross sections as low as $\bar{\sigma}_e \lesssim 10^{-40}~\mathrm{cm}^2$ with gram-scale exposures. We note that there is some uncertainty for $\omega \gtrsim 15$~eV ($q \gtrsim 20$ keV) where ELF models diverge.

\subsection{Absorption} \label{sec:absorption}

Light bosonic DM particles with masses in the sub-keV range can be efficiently absorbed by a material, in analogy with photon absorption. 
Let us consider a dark photon $A_{\mu}'$ with mass $m_{V}$ kinetically mixed with the photon\footnote{
    See Refs.~\cite{Graham:2015rva,Ema:2019yrd,Long:2019lwl,Kitajima:2022lre,Nakayama:2019rhg,Nakayama:2020rka,Kitajima:2023fun,Agrawal:2018vin,Co:2018lka,Bastero-Gil:2018uel,Dror:2018pdh,Nakayama:2021avl} for production mechanisms of the dark photon DM.
}
\begin{align}
\mathcal{L} \supset 
 -\frac{1}{4} F^{\mu\nu}F_{\mu\nu}
- \frac{1}{4} F'^{\mu\nu}F'_{\mu\nu}
- \frac{\kappa}{2} F^{\mu\nu}F'_{\mu\nu}
+ \frac{1}{2} m_{V}^2 A'_{\mu}A'^{\mu},
\end{align}
where $\kappa$ is the kinetic mixing parameter.
After removing the kinetic mixing term with the field redefinition, the dark photon couples to electric current with a suppressed coupling $\kappa e$. 
The coupling to the electromagnetic current enables absorption into electronic or phononic excitations. 
The dark photon mass $m_V$ determines which absorption process is relevant, since the absorbed energy $\omega$ is almost equal to $m_V$.

\begin{figure*}[t]
    \centering
    \includegraphics[width=0.475\linewidth]{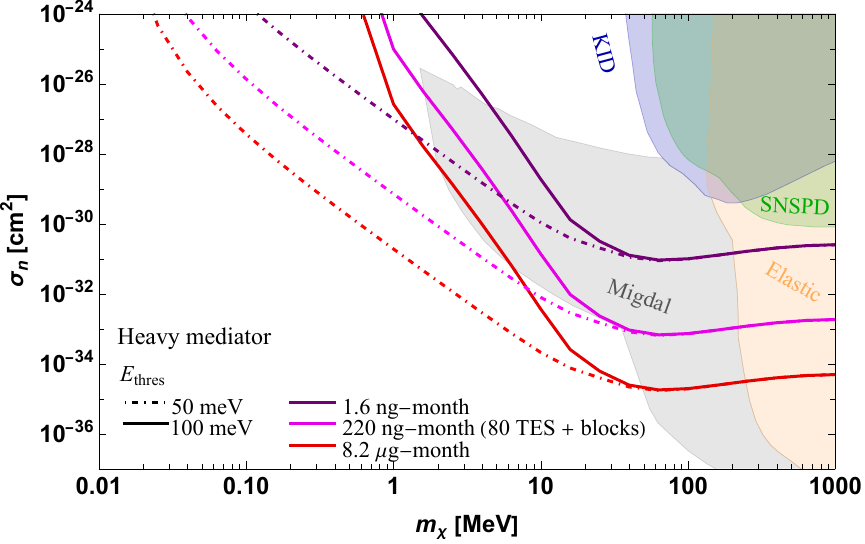}
 \hspace{1em}
 \includegraphics[width=0.475\linewidth]{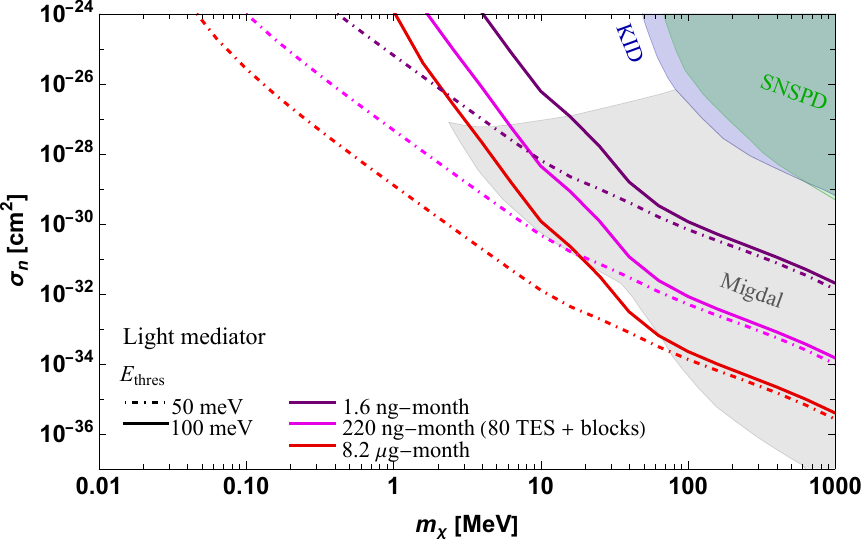}
        \caption{[Left] Sensitivity projections for TES detection of DM-nucleon scattering with heavy mediator at the $95~\%$ C.L., for various exposures (solid lines). The blue and green shaded region indicates bounds derived from SNSPD data~\cite{Hochberg:2019cyy,Hochberg:2021yud} and KID data~\cite{Gao:2024irf}, respectively. The orange shaded region and the gray shaded region indicate the existing constraints based on elastic collisions and Migdal effect, respectively, including XENON1T~\cite{PhysRevLett.124.021801}, PANDA-4T~\cite{PhysRevLett.131.191002}, LUX~\cite{PhysRevLett.122.131301}, CRESST-III~\cite{PhysRevD.100.102002}, DarkSide-50~\cite{PhysRevD.107.063001,franco2023lightdarkmattersearch}, EDELWEISS~\cite{PhysRevD.99.082003,PhysRevD.106.062004}, SuperCDMS~\cite{PhysRevLett.127.061801,PhysRevD.107.112013}, and SENSEI~\cite{Adari_2025}. [Right] DM-nucleon scattering for light mediator.}
    \label{fig:DMnucleon}
\end{figure*}

The absorption rate per unit detector mass is governed by the imaginary part of the dielectric function in the $q \to 0$ limit 
(see App.~\ref{app:calc} for more details)
\begin{equation}
\label{eq:Rabs}
R = \frac{\kappa^2 \rho_{\chi}}{\rho_T}   \mathrm{Im} \left[ -\frac{1}{\epsilon(m_{V})} \right],
\end{equation}
where $\epsilon(\omega)$ is the complex dielectric function evaluated at $\omega = m_{V}$. 

DM absorption can proceeds through excitation of conduction band electrons. At lower energies ($m_{V} \lesssim 0.1~\mathrm{eV}$), the dominant absorption channels is multiphonon process - there are no optical phonon modes for Al. The ELF in this regime is well measured or modeled using analytical fits~\cite{Knapen:2021bwg}. We extract the ELF for Al from optical data using the Mermin model, and evaluate $\mathrm{Im}[-1/\epsilon(\omega)]$ in the $q \to 0$ limit using \texttt{DarkELF}.  

Fig.~\ref{fig:DMabsorption} shows projected sensitivities for dark photon DM absorption in our TES setup, assuming benchmark exposures. 
A clear plasmon peak is seen at $m_V \simeq 15$ eV. 
A dip around $\sim 70$ eV corresponds to the $L$ edge~\cite{PhysRevB.22.1612,HENKE1993181}.
With a 0.1 eV threshold and no excess background, the single-photon sensitivity enables testing kinetic mixing down to $\kappa \sim 10^{-15}$, surpassing astrophysical limits in the meV-eV mass range.

\subsection{Nucleon scattering} \label{sec:dmn}

In addition to DM-electron scattering, our TES setup is sensitive to interactions where sub-GeV DM scatters coherently off target nuclei. 
This is the case if the mediator interacts with nucleon $n$ through $\mathcal L =  g_n A_\mu' J^\mu_n$ in Eq.~(\ref{eq:L_scat_e}).
The associated scattering rate is governed by the nuclear structure factor $S_n(q, \omega)$.
Note that, if the mediator only couples to proton (but not to neutron), for example, then $S_n(q, \omega)$ should be multiplied by $(Z/A)^2$ with $Z$ $(A)$ being the atomic (mass) number of the nucleus.

The differential scattering rate is calculated in a similar way to Eq.~(\ref{eq:dRdomega_e}) as
\begin{align}
\label{eq:dRdomega_n}
    \frac{dR}{d\omega} =&~ \frac{\rho_\chi}{\rho_T m_\chi} \frac{\bar{\sigma}_n}{\mu_{\chi n}^2}
    \int_{q_-}^{q_+} \frac{qdq}{4\pi} \left|F_{\rm med}(q)\right|^2 \nonumber\\
    &~\times S_n(\omega, q) \eta\left(v_{\rm min}(\omega, q)\right)~,
\end{align}
where $F_{\rm med}(q)$ takes the same form given in Eq.~\eqref{eq:Fmed} but with a different reference momentum transfer $q_{0,n}=m_\chi v_0$. Here,
$\bar{\sigma}_n$ is the reference DM--nucleon cross section given by
\begin{eqnarray}   \bar{\sigma}_n=\frac{g_n^2g_\chi^2\mu_{\chi n}^2}{\pi(m_\chi^2 +q_{0,n}^2)^2}
\end{eqnarray}
with the DM–nucleon reduced mass $\mu_{\chi n} = m_{\chi} m_n/(m_{\chi} + m_n)$.
The total rate is then obtained by Eq.~\eqref{eq:dRdo_to_R}.

We consider the full structure factor $S_n(q,\omega)$, comprising both single-phonon contributions at low $q$ momentum transfer and multiphonon or impulse-approximation  tails at higher $q$. This unified approach enables a comprehensive treatment of DM-nucleus scattering in materials with crystalline structure.
However, the phonon regime, which dominates the low-energy nuclear response, is not accessible in our TES setup due to the limited detection threshold. Specifically, phonon excitations lie in the $\mathcal{O}($meV) range, while our TES has an effective threshold of $E_{\rm th} = 0.1$~eV. As a result, for our projected exposures and threshold, phonon and multiphonon excitations are not observable, and the rate is entirely dominated by nuclear recoil tails at higher $\omega$, consistent with a standard elastic recoil treatment.

In Fig.~\ref{fig:DMphonon} we illustrate sensitivity for DM-nucleon scattering with heavy mediator considering different energy thresholds, showcasing significant suppression of phonon-induced scattering at $E_{\rm th} \gtrsim 0.1~\mathrm{eV}$. For DM masses $m_\chi \lesssim 100~\mathrm{MeV}$, most scattering would proceed via phonon channels. However, these contributions become suppressed above threshold, leading to rapidly decreasing sensitivity. 

\begin{figure*}[t]
    \centering
    \includegraphics[width=0.475\linewidth]{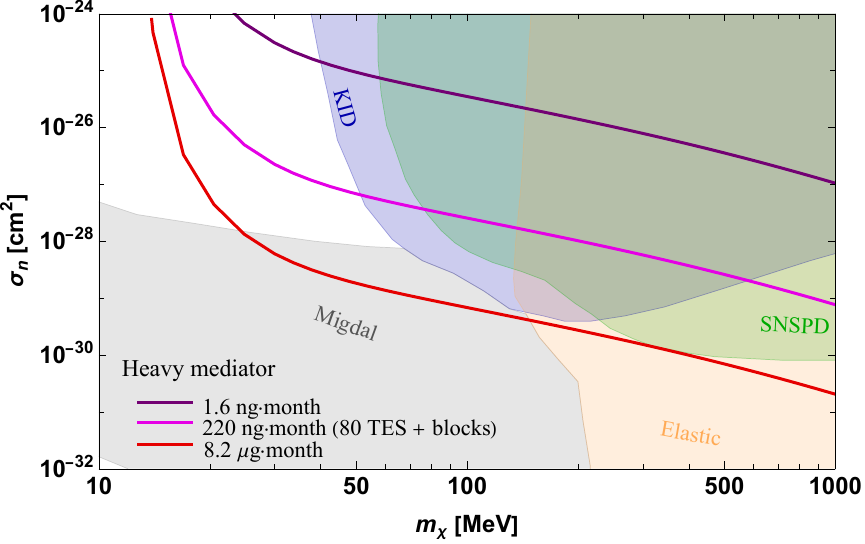} \hspace{1em}
    \centering
    \includegraphics[width=0.475\linewidth]{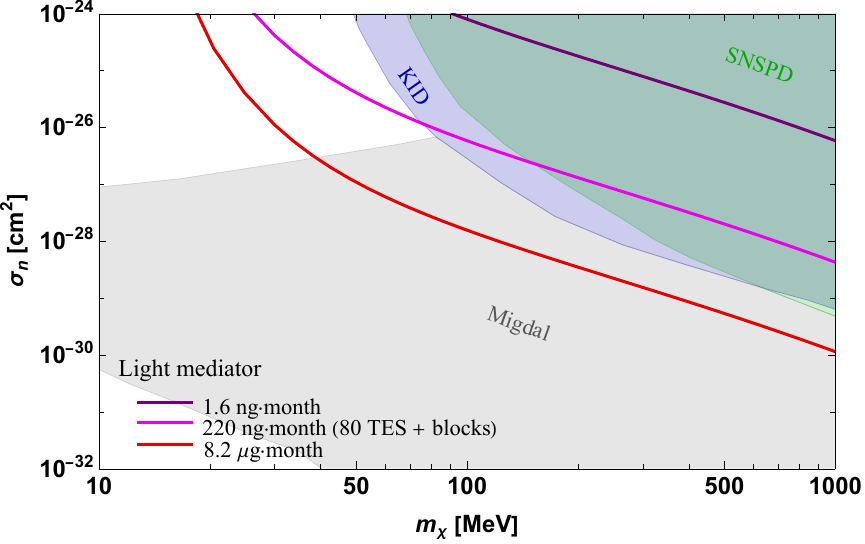}
    \caption{[Left] Sensitivity projects for TES detection of DM-nucleon scattering through the Migdal effect with heavy mediator at the $95~\%$ C.L., considering different exposures (solid lines). The solid lines are calculated using the free atom approximation with nuclear recoil threshold $E^{\rm thr}_N=0.1~{\rm eV}$ and electronic excitation threshold $\omega_{\rm th}=5 {\rm ~eV}$. Here we also show the same existing constraints shown in Fig.~\ref{fig:DMnucleon}.
    [Right] DM-nucleon scattering through the Migdal effect with light mediator.}
    \label{fig:migdal}
\end{figure*}

In Fig.~\ref{fig:DMnucleon} we further demonstrate that in both the light and heavy mediator  scenarios, respectively, sensitivity improves substantially with lower thresholds.
The regime transitions, from phonon-dominated to impulse-dominated, depend on the energy threshold. Our scenarios lie in the latter.

\subsection{Migdal effect} \label{sec:migdal}

We distinguish between elastic DM–nucleon scattering and the Migdal effect analysis. The Migdal effect~\cite{Migdal:1939,Migdal:1941} refers to an inelastic atomic process wherein a DM–induced nuclear recoil causes a sudden displacement of the nucleus relative to the electron cloud, leading to the ionization or excitation of bound electrons~\cite{Bernabei:2007jz,Ibe:2017yqa,Knapen:2020aky}. This effect provides additional electromagnetic signatures that could potentially be more easily detectable than the recoils itself. These electromagnetic signatures are often easier to detect than the nuclear recoils themselves.  While the elastic channel discussed in the previous subsection produces nuclear recoils or at lower energies lattice phonons, the Migdal channel yields prompt electronic energy deposits that can be recorded calorimetrically by TES.
 
For an isotropic target material such as Al, the differential event rate is
\begin{eqnarray}
    \frac{dR}{d\omega dE_N}\simeq \frac{\rho_\chi}{m_N m_\chi}\int_{v_{\rm min}} d^3v  vf(v)\frac{d\sigma}{dE_N}\frac{dP}{d\omega},
\end{eqnarray}
where $m_N$ is the mass of Al nucleus, $E_N = q^2/2m_N$ is the nuclear recoil energy and $v_{\rm min}=\sqrt{2E_N/\mu_{\chi N}}$ is the minimum velocity for recoil. The probability density for Migdal transitions considering deposited energy $\omega$ in electronic system
\begin{eqnarray}
    \frac{dP}{d\omega}=\frac{4\alpha_e Z^2_{\rm ion}E_N}{3\pi^2 \omega^4m_N}\int dk k^2{\rm Im}\left[\frac{-1}{\epsilon(k,\omega)}\right]~.
\end{eqnarray}

The factor $1/E_N(\partial P/\partial\omega)$ is independent of $v$ and $E_N$. Hence, following Ref.~\cite{Knapen:2021bwg}, we can define
\begin{eqnarray}
\label{eq:J}
    J(v,\omega)=\int dE_N E_N\frac{d\sigma}{dE_N}
\end{eqnarray}
so that differential rate with respect to $\omega$ becomes
\begin{eqnarray}
    \frac{dR}{d\omega}\simeq\frac{\rho_\chi}{m_\chi m_N}\left[\frac{1}{E_N}\frac{\partial P}{\partial \omega}\right]\int_{v_{\rm min}}d^3v vf(v)J(v,\omega)~.
\end{eqnarray}
Within the free atom approximation
in the massive mediator $(m_V\rightarrow \infty)$ and massless mediator $(m_V\rightarrow 0)$ limits, Eq.~\eqref{eq:J} reduces to~\cite{Knapen:2021bwg} 
\begin{eqnarray}
    J_{\infty}&=&\frac{A^2\bar{\sigma}_n}{16\pi^2m_N v^2\mu_{\chi n}^2}(q_+^4-q^4_-)\\
    J_0&=&\frac{A^2\bar{\sigma}_n q_{0,n}^4}{4\pi^2m_N v^2\mu_{\chi n}^2}\log\left[\frac{q_+}{q_-}\right]~.\notag   
\end{eqnarray}
The momentum transfer limits are taken as
\begin{align}
q_{+}=&~v \mu_{\chi N} \left(1+\sqrt{1-\frac{2\omega}{v^{2}\mu_{\chi N}}}\right),\\
q_{-}=&~\max \left[
      v \mu_{\chi N} \left(1-\sqrt{1-\frac{2\omega}{v^{2}\mu_{\chi N}}}\right),
      \sqrt{2 m_{N} E_{N}^{\mathrm{thr}}}
\right] \notag
\end{align}
with $E_{N}^{\mathrm{thr}}$ is a possible nuclear recoil-energy analysis cut.
For our calorimetric TES analysis we set here $E_{N}^{\rm thr}= 0.1$~eV.

Al is a metal and has no band gap. Promoting an electron above the Fermi sea requires overcoming the work function $\simeq4.1~\text{eV}$. 
Hence, for Migdal channel we conservatively integrate electronic excitations above these energies with
 $\omega\gtrsim 5~\mathrm{eV} $. 
In semiconductors, such as silicon and germanium, the analogous energy threshold
can instead be set approximately by the energy required to generate two ionization
electrons $2e^{-}$~\cite{Essig:2015cda}. 
The TES energy threshold $E_{\rm thr} = 0.1$~eV is applied to the total deposited energy $E_{\rm dep} = E_N + \omega$ and is far below the considered $\omega$ energies. Hence, TES threshold in our analysis does not limit Migdal effect reach.

In Fig.~\ref{fig:migdal} we display the resulting DM-nucleon cross section sensitivity for heavy and light mediators, respectively. We observe that our TES setup can achieve for multi-pixel configuration sensitivity that exceeds existing SNSPD and KID limits.

\subsection{Power deposition}

DM can also be probed by quantum sensors through power deposition~\cite{Das:2022srn}. The total power deposited in the TES absorber due to DM-nucleon scattering is 
\begin{eqnarray}
    P  =   \int_\Delta d\omega   \omega\frac{dR}{d\omega},
\end{eqnarray}
where we consider contributions above the superconducting Al gap energy $\Delta=340~{\rm \mu eV}$. The upper $\omega$ integration limit is determined by vanishing $S(q,\omega)$, typically $\mathcal{O}(100~{\rm meV})$. 

For characteristic values of $m_\chi\sim100$~MeV  and $\bar{\sigma}_n=10^{-32}~{\rm cm^{2}}$, from Eq.~\eqref{eq:dRdomega_n} the maximum rate is $dR/d\omega \sim 10^{-8}~({\rm meV \mu g  s})^{-1}$. Hence, the resulting deposited power can be roughly estimated, assuming $\omega \simeq d\omega \simeq 0.1$~eV, as $P \sim 10^{-7}~{\rm eV~(\mu g  s)^{-1}}$, or equivalently $P \sim 10^{-31}~{\rm W \mu m^{-3}}$. Considering as reference optical TES realized in Ref.~\cite{Hattori:2022mze} with a bias power of $ 1.4\times10^{-13}$ W and Al absorber with a  volume of $10^3$ $\mu \text{m}^3$, i.e. single pixel with typical size 100 $\mu$m $\times$ 100 $\mu$m $\times$ 0.1 $\mu$m, yields a power density of $P_{\rm TES} \simeq 1.4\times 10^{-16}$ W$\mu \text{m}^{-3}$. Thus, even under optimistic considerations, the DM-induced power  $P  $ can be expected to be orders of magnitude below the TES detection threshold.

\section{Conclusions}
\label{sec:conc}

We have demonstrated that multiplexed optical TES arrays operating as quantum sensors near thermodynamic noise limit capable of resolving sub-eV energy deposits can significantly broaden the experimental frontier of sub-GeV DM. Our analysis shows that state-of-the-art high resolution optical TES, recently experimentally demonstrated with energy resolution below $\lesssim 70~\mathrm{meV}$ and thresholds around $0.1$~eV, offer unprecedented sensitivity to DM interactions. We comprehensively model   realistic TES noise and pulse response as well as in-medium material effects for DM interactions. We demonstrate that the same TES platform can simultaneously efficiently target multiple DM interactions--including DM-electron scattering, DM absorption, and DM-nucleon scattering and DM-Migdal-assisted nuclear recoils-- making it particularly versatile among quantum sensors. We show that even ng-month experimental exposures can probe DM–electron cross sections down to $\sigma_e \lesssim 10^{-27}~\mathrm{cm}^2$ for sub-MeV DM masses, and DM–nucleon cross-sections down to $\sigma_n \lesssim 10^{-26}~\mathrm{cm}^2$ for DM masses at the MeV scale. A scalable TES array with 80 pixels can improve this reach by approximately two orders of magnitude, with further gains achievable through additional multiplexing.

Given sub-eV energy resolution, photon-number discrimination capabilities, and potential for low-background operation in underground environments, high-resolution TES arrays offer a compelling and scalable quantum sensing technology for future rare-event searches and DM.

\medskip\noindent\textit{Acknowledgment.---} 
We would like to thank Suerfu Burkhant, Maurice Garcia-Sciveres, Kazuhisa Mitsuda, Tien-Tien Yu, Yu Zhou for valuable discussions. V.T. would like to thank Fermilab Theory Division for hospitality where part of this work was conducted. This work was supported by World Premier International Research Center Initiative (WPI), MEXT, Japan. We acknowledge support by the JSPS KAKENHI grant No. 23K13109 (V.T.,) No. 24K07010 (K.N.) and No. 25H02197 (K.H.), and JST-FOREST Program Grant No. JPMJFR2236.

\appendix

\section{Theoretical Framework}
\label{app:calc}

\subsection{Intrinsic material properties}

The space-time correlation function of an operator $\mathcal{O}(t, {\bf x})$, typically representing the electron or nuclear number density in a material, is defined as
\begin{align}
    S_{\mathcal O} (\omega, {\bf x}) = \frac{1}{V_T} \int dt \int d^3x' e^{i\omega t}
    \left< \mathcal O(t,{\bf x + x'}) \mathcal O(0,{\bf x'}) \right>,
\end{align}
where $V_T$ is the material volume and the expectation value is taken over thermal equilibrium state ensemble  $\left<\bullet\right> = \sum_{m}p_m \left<m|\bullet |m\right>$, where $p_m = e^{-E_m/T}/Z$ and $Z$ being the partition function.
 The dynamic structure factor, which characterizes the material response to external probes, is obtained by taking the Fourier transform in space
\begin{align}
    S_{\mathcal O} (\omega, q) &= \int d^3x e^{-i {\bf q\cdot x}} S_{\mathcal O} (\omega, {\bf x}) \\
   &= \frac{1}{V_T} \int dt e^{i\omega t}
    \left< \mathcal O(t,{\bf q}) \mathcal O(0,{\bf -q}) \right>,
\end{align}
where the operator in momentum space is defined by 
\begin{align}
    \mathcal{O}(t, {\bf x}) = \int \frac{d^3q}{(2\pi)^3} e^{i {\bf q \cdot x}} \mathcal{O}(t, {\bf q}).
\end{align}
The dynamic structure factor admits a K\"{a}llen–Lehmann spectral representation 
\begin{align}
    \label{eq:SO}
    S_{\mathcal O} (\omega, q) = \frac{2\pi}{V_T}\sum_{m,n} p_m
    \left|\left<m| \mathcal O({\bf q})|n\right>\right|^2  \delta(\omega+E_m-E_n).
\end{align}
This quantity encapsulates intrinsic properties of the medium and governs its response to weak external perturbations such as DM interactions.

To understand how a material responds to an external perturbation—such as an electromagnetic wave or DM interaction—the (retarded) susceptibility $\chi_{\mathcal{B}\mathcal{A}}(\omega, q)$ plays a central role. It quantifies the linear response of the expectation value of the operator $\mathcal{B}$ due to a perturbation coupled to the operator $\mathcal{A}$, and is defined by
\begin{align}
     \chi_{\mathcal BA} (\omega, q) 
    =&~ \frac{1}{V_T}\int_0^\infty dt e^{i\omega t} i \left< [ \mathcal B(t,{\bf q}), \mathcal A(0,-{\bf q})] \right> \\    
    =&~- \frac{1}{V_T}\sum_{m,n} \frac{(p_m-p_n)\left<m| \mathcal A(-{\bf q})|n\right>\left<n| \mathcal B({\bf q})|m\right> }{\omega+E_m-E_n + i\eta}, \notag
\end{align}
where $p_m = e^{-E_m/T}/Z$ is the Boltzmann weight, and $\eta \to 0^+$ is taken at the end of the calculation to ensure causal (retarded) behavior. 

For the case where $\mathcal{B}^\dagger = \mathcal{A} = \mathcal{O}$, and using the identity
\begin{align}
    \lim_{\eta\to 0} \frac{1}{x+i\eta} = {\rm P}\left(\frac{1}{x}\right)-i\pi\delta(x),
\end{align}
we obtain the imaginary part of the susceptibility,
\begin{align}
    {\rm Im} \chi_{\mathcal O^\dagger O} (\omega, q) 
    =&~ \frac{\pi(1-e^{-\omega/T})}{V_T}\sum_{m,n} p_m
    \left|\left<m| \mathcal O({\bf q})|n\right>\right|^2 \nonumber\\
    &~\times (2\pi) \delta(\omega+E_m-E_n).
\end{align}
The imaginary part of $\chi$ encodes the dissipative response of the system to external driving at frequency $\omega$ and momentum $q$.

This leads directly to the fluctuation–dissipation theorem,
\begin{align}
    {\rm Im} \chi_{\mathcal O^\dagger O} (\omega, q) 
    = \frac{1-e^{-\omega/T}}{2} S_{\mathcal O} (\omega, q).
\end{align}
which relates the dissipative response to the intrinsic thermal fluctuations of the operator $\mathcal{O}$. In particular, when $\mathcal{O}$ corresponds to the electron number density, $\chi$ is proportional to the photon self-energy polarization tensor\footnote{Up to a factor of the electromagnetic coupling $e^2$.}, and thus directly linked to the material's electric conductivity and ELF. The latter is given by $\text{Im}[-1/\epsilon(\omega, q)]$, as expressed in Eq.~\eqref{eq:Sqw_eps}.

\subsection{DM scattering}

As an example, let us consider the interaction Hamiltonian density
\begin{align}
    \mathcal H_1 = \phi (g_\chi n_\chi + g_i n_i),  \label{H1_scat}
\end{align}
where $\phi$ is a mediator scalar, $n_\chi$ and $n_i$ are number density operators of DM particle $\chi$ and SM field $i$ (electron or nucleon), respectively\footnote{
    The operator of the form $\bar\chi\chi$ corresponds to number density in the non-relativistic limit. In the case of vector mediator $A'_\mu$, it couples to the current $\bar\chi\gamma^\mu\chi$ but it reduces to the form of Eq.~\eqref{H1_scat} since $A_0'$ exchange dominates the interaction.
}.
The initial state $\left| \psi_I(0) \right>$ in the interaction picture evolves according to
\begin{align}
    \left| \psi_I(t)\right> = {\rm T}\exp\left[-i \int_0^t H_I(t) \right] \left| \psi_I(0)\right>,
\end{align}
where $\mathrm{T}$ denotes time ordering, and
\begin{align}
    H_I(t) = e^{iH_0 t}\left(\int d^3x \mathcal H_1(x)\right)e^{-iH_0 t}.
\end{align}

Suppose the initial state is $\left|p_\chi; i\right>$ and the final state is $\left|p'_\chi; f\right>$, where $p_\chi$ and $p'_\chi$ denote the momenta of the initial and final dark matter particle $\chi$, and $i$ and $f$ represent the initial and final states of an electron in the material.
The transition amplitude arises at second order in perturbation theory with respect to $H_I$ and is given by
\begin{align}
    \alpha_{fi}(t) =&~ -g_\chi g_i \left<p_\chi';f\right| \int d^4x_1 \int d^4x_2  \\
   &\times {\rm T} \left[ n_{\chi,I} (x_1) \phi_I(x_1)\phi_I(x_2) n_{i,I}(x_2) \right]
   \left|p_\chi; i\right>\notag\\
    =&~-g_\chi g_i   \int d^4x_1 \int d^4x_2\int\frac{d^4q}{(2\pi)^4}\frac{i}{q_\mu^2-m_\phi^2}\nonumber\\
   & \times e^{i(E_\chi-E_\chi'-\omega)t} e^{i(E_i-E_f+\omega)t} \nonumber\\
   & \times  \left<p'_\chi\right| n_\chi(x_1)e^{i{}\bf q\cdot x_1} \left| p_\chi\right>
    \left<f\right| n_i(x_2)e^{-i\bf q\cdot x_2} \left|i\right>, \notag
\end{align}
where we have defined the four-momentum $q^\mu = (\omega, \mathbf{q})$.

Making use of the Fourier-transformed operators for $n_i$ and $n_\chi$, the amplitude can be rewritten as
\begin{align}
   \alpha_{fi}(t) &= -g_\chi g_i  \int_0^t dt e^{i(E_f-E_i-\omega)}\int\frac{d^3q}{(2\pi)^3}\frac{i}{q_\mu^2-m_\phi^2}\nonumber\\
    &~~~\times \left<p'_\chi\right| n_\chi(-q) \left| p_\chi\right>
    \left<f\right| n_i(q) \left|i\right>.
\end{align} 
Here the momentum eigenstate of the DM particle is defined as
\begin{align}
    \left|p_\chi\right> = \frac{a^\dagger_{p_\chi}}{\sqrt V} \left|0\right>,
\end{align}
with $a^\dagger_{p_\chi}$ the creation operator and $V$ the quantization volume, taken to infinity at the end of the calculation.
The transition probability per unit time is then given by
\begin{align}
   \Gamma_{fi} &= \frac{|\alpha_{fi}(t)|^2}{t} \nonumber\\
   &= \left( \frac{g_\chi g_i}{V(q_\mu^2-m_\phi^2)}\right)^2 \left| \left<f\right| n_i(q) \left|i\right>\right|^2 \nonumber\\
   &~~~~~~\times (2\pi)\delta(E_i-E_f-\omega).
\end{align}

This result can be interpreted as Fermi's golden rule. It represents the scattering rate for the case where a single $\chi$ particle with momentum $p_\chi$ in a volume $V$ transitions to a final momentum $p_\chi'$. To obtain the physical scattering rate, we must multiply by the total number of DM particles in the volume, $N\chi = \rho_\chi V / m_\chi$, and take an average over the DM velocity distribution $f(v)$.
In addition, we integrate over the allowed final states, which contributes a factor $V \int d^3q/(2\pi)^3$, and perform a thermal average over the initial electron states $\left| i \right>$ in the material. This yields the total scattering rate as
\begin{align}
   \Gamma
   =&~\frac{\rho_\chi}{m_\chi} \int d^3 vf(v) \int\frac{d^3q}{(2\pi)^3} \left( \frac{g_\chi g_i}{q_\mu^2-m_\phi^2}\right)^2 \nonumber\\
   &\times\sum_{i,f}p_i\left| \left<f\right| n_i(q) \left|i\right>\right|^2 (2\pi)\delta(E_i-E_f-\omega).
\end{align}
where $p_i$ is the thermal occupation probability of the initial state $\left| i \right>$.
Identifying in above the form of the dynamic structure factor $S_i(\omega, q)$ defined in Eq.~\eqref{eq:SO}, we  obtain Eq.~\eqref{eq:SO}
\begin{align}
   \Gamma
   =\frac{\rho_\chi V_T}{m_\chi} \int d^3 vf(v) \int\frac{d^3q}{(2\pi)^3} \left( \frac{g_\chi g_i}{q_\mu^2-m_\phi^2}\right)^2 S_i (\omega, q).
\end{align}
where $V_T$ denotes the total target volume.

To obtain the scattering rate per unit detector mass, we divide the total rate by the total target mass $M_T = \rho_T V_T$, yielding
\begin{align}
   R 
   =\frac{\rho_\chi}{\rho_T m_\chi} \int d^3 vf(v) \int\frac{d^3q}{(2\pi)^3} \left( \frac{g_\chi g_i}{q_\mu^2-m_\phi^2}\right)^2 S_i (\omega, q).
\end{align}
This expression is used in Eq.~\eqref{eq:R_scatter}, after making the standard approximation $q^\mu q_\mu \simeq -\mathbf{q}^2$, which is valid in the non-relativistic regime.
Here, $\omega$ should be regarded as a function of $q$ and $v$, determined by the kinematic relation given in Eq.~\eqref{eq:DeltaE}.

\subsection{DM absorption}

Next, let us consider the absorption of dark matter by the material. Focusing for illustration on dark photon DM, the relevant interaction Hamiltonian density is
\begin{align}
   \mathcal H_1 = \kappa e A_\mu' J_e^\mu.
\end{align}
where $\kappa$ is the kinetic mixing parameter, $A'_\mu$ is the dark photon field, and $J_e^\mu$ is the electromagnetic current of the electrons.

In this case, the transition occurs at first order in perturbation theory with respect to $H_I$. Repeating a similar procedure as in the previous subsection, we find the transition amplitude
\begin{align}
   \alpha_{fi}(t) 
    =&~ \left<f\right|\left[-i\int_0^t dt H_I(t)\right] \left|p_A; i\right> \notag\\
    =&~-i\kappa e \int_0^t dt e^{i(E_f-E_i-\omega)}
   \int\frac{d^3q}{(2\pi)^3} \nonumber \\
   & \times \left<0\right| A_\mu'(q) \left|p_A\right>\left<f\right| J^\mu(-q) \left|i\right>,
\end{align}
for a one-particle initial dark photon state.

The total absorption rate is then given by
\begin{align}
   \Gamma &= \frac{\kappa^2e^2 \rho_V}{2\omega^2} \sum_{i,f} \left|\epsilon_\mu \left<f\right|J_e^\mu(-q)\left|i\right> \right|^2
   \nonumber\\
   &~~~\times (2\pi)\delta(E_f-E_i-\omega).
\end{align}
where $\epsilon_\mu$ is the polarization vector of the dark photon, and $\rho_V$ is the local dark photon DM energy density.
For a massive, non-relativistic dark photon, the time-like ($\mu = 0$) component is suppressed and can be safely neglected. Since the two transverse and one longitudinal modes contribute similarly, we average over the polarizations. Using current conservation $q_\mu J^\mu = 0$, we arrive at
\begin{align}
   \Gamma = \frac{\kappa^2e^2 \rho_V V_T}{2q^2} S_e(\omega, q),
\end{align}
where $\omega \simeq m_V$ and $q \simeq 0$ in the non-relativistic limit.

The absorption rate per unit target mass is thus
\begin{align}
   R &= \frac{\kappa^2e^2 \rho_V}{2q^2 \rho_T} S_e(\omega, q) \nonumber\\
   &= \frac{\kappa^2 \rho_V}{\rho_T} {\rm Im} \left[\frac{-1}{\epsilon(\omega, q)}\right],
\end{align}
where we have used\footnote{Note that the absorption of scalar dark matter cannot be accurately captured by the dielectric function formalism~\cite{Mitridate:2021ctr}.} Eq.~\eqref{eq:Sqw_eps}. This yields Eq.~\eqref{eq:Rabs}.  

\bibliography{references}

\end{document}